\documentclass[pra,twocolumn,superscriptaddress]{revtex4-2}
\usepackage{graphicx,amsmath,amssymb,hyperref} 
\usepackage{caption,color}
\usepackage{subcaption}
\usepackage{comment}

\begin{document}
\title{Single and coupled quantum van der Pol oscillators: Wigner-function dynamics, exact qutrit solutions, and synchronization}

\author{Rohitashwa Chattopadhyay}
\thanks{These authors contributed equally to this work.\\
Corresponding author: Vinayak Pendse
(vinayak.pendse@uni-saarland.de).}
\affiliation{Department of Theoretical Physics, Tata Institute of Fundamental Research, Homi Bhabha Road, Colaba, Mumbai 400005, India}

\author{Vinayak Pendse}
\thanks{These authors contributed equally to this work.\\
Corresponding author: Vinayak Pendse
(vinayak.pendse@uni-saarland.de).}
\affiliation{Indian Institute of Science Education and Research,
Dr. Homi Bhabha Road, Pashan, Pune 411008, India; Current affiliation: Theoretische Physik, Universit\"{a}t des Saarlandes, D-66123 Saarbr\"{u}cken, Germany}

\affiliation{Center for Quantum Technologies (QuTe), Saarland University, Campus, 66123 Saarbr\"ucken, Germany}

\author{Shamik Gupta}
\affiliation{Department of Theoretical Physics, Tata Institute of Fundamental Research, Homi Bhabha Road, Colaba, Mumbai 400005, India}

\date{\today}
\begin{abstract}
We study a quantum van der Pol oscillator in which single-photon pumping competes with a nonlinear two-photon dissipative channel to produce a quantum limit cycle, and the question of how two such oscillators with non-identical intrinsic frequencies synchronize in the sense of phase locking when coupled dissipatively to one another. For the single oscillator, starting from a Lindblad description of the dynamics, we derive the exact Wigner-function evolution equation, and show that its
Fokker--Planck approximation admits in the appropriate
regime an equivalent classical It\^{o} stochastic
description that reproduces remarkably well the full Lindblad dynamics. Notably, the stochastic approach provides an efficient means of capturing system dynamics in the semiclassical regime, where a large number of energy eigenstates contribute and direct numerical simulations  become computationally demanding. Restricting to the three lowest Fock states and using a parity symmetry, we obtain in closed form the stationary-state density matrix and its Wigner-function representation that displays pronounced nonclassical features. For
two non-identical oscillators, we study the full Lindblad dynamics
numerically and characterize synchronization through the stationary-state
order parameter as a function of frequency mismatch and dissipative
coupling. We find that increasing the effective nonlinear dissipative
strength suppresses the synchronized region. Restricting the two
oscillators to the three-level qutrit subspace, a parity symmetry reduces
the stationary-state problem to an exact $41$-dimensional linear system,
allowing to compute the synchronization order parameter.
In this reduced description, results in the limit of large dissipation agree closely with those from the full dynamics, while exhibiting in the limit of small dissipation a
double-peak synchronization structure accompanied by a phase reversal.
Our results provide an analytical framework for understanding quantum
synchronization in dissipatively-coupled self-sustained oscillators.   
\end{abstract}
\maketitle

\tableofcontents

\section{Introduction}
Self-sustained oscillators are systems that maintain persistent oscillations through a balance between energy gain and loss, without the application of any periodic external drive. Limit-cycle oscillators are a specific class of self-sustained oscillators whose long-time dynamics approaches a stable, isolated periodic orbit in phase space, making the oscillations robust against small perturbations. Such oscillators are ubiquitous in nature across physical, chemical, biological, and even artificial systems~\cite{strogatznonlinear}. Two or more such oscillators when interacting with one another can adjust their rhythms and lock to a common frequency and phase, a phenomenon commonly known as synchronization~\cite{PRK:2001,Boccaletti2002,Gupta2018}. First observed by Huygens in the seventeenth century, the phenomenon of synchronization has since become a paradigm of collective behavior in nonlinear dynamics and statistical physics. 

A central model in this context that has over the years played a crucial role in the development of the subject is the van der Pol oscillator, in which linear negative damping and nonlinear positive damping combine to produce a stable limit cycle in phase space~\cite{KOVACIC2026120004}. The celebrated Kuramoto model provides a minimal framework to describe how a system of limit-cycle oscillators with distributed frequencies when interacting with one another evolves to a state exhibiting collective synchronization~\cite{Kuramoto1984,Acebron2005,gupta2014kuramoto,rodrigues2016kuramoto}. Synchronization can refer to frequency locking, phase locking, or both. In frequency synchronization, oscillators adjust their long-time average frequencies to a common value, even if their phases continue to drift relative to one another. By contrast, in phase synchronization, the phase difference between the oscillators remains bounded in time; In the stronger case of phase locking, the difference approaches a time-independent value. Strict phase locking implies frequency locking, but the converse is not necessarily true.

Extending the ideas of synchronization to the quantum domain has become an active research direction in recent years, motivated by advances in controlling small, strongly-dissipative systems such as trapped ions, optomechanical devices, and superconducting circuits~\cite{solanki2026quantumsynchronization,lahiri2026quantumsynchronizationtutorialguide}. Quantum synchronization has been studied in a range of settings, including quantum van der Pol oscillators, spin systems, and networks of oscillators~\cite{nadolny2026quantum,kehrer2024quantum,kehrer2025quantum,nadolny2025nonreciprocal,solanki2024exotic,nadolny2023macroscopic,daniel2023geometric,koppenhoefer2019optimal,koppenhoefer2020quantum,roulet2018quantum,roulet2018synchronizing,lorch2017quantum,lorch2016genuine,walter2015quantum,walter2014quantum,Vigneshwar2025,Vigneshwar2026,Vigneshwar2026b,Bandyopadhyay2023,Paul2024,Bandyopadhyay2021,Preethi2025,Thomas2026,solanki2026quantumsynchronization}. The study of synchronization in the quantum regime raises a number of subtleties that have no classical counterpart: Fluctuations are intrinsic, the phase-space description via Wigner function, etc. is only quasi-probabilistic, and the notions of a limit cycle and of phase locking must be suitably redefined in terms of density operators and expectation values. A convenient formulation of the quantum van der Pol oscillator was proposed in Ref.~\cite{Arosh2021} by Arosh \textit{et al.}, which combines single-photon pumping with a nonlinear two-photon dissipative channel in the framework of a Lindblad master equation for an open quantum harmonic oscillator. The system in its stationary state displays a quantum limit cycle in the Wigner representation. In this work, we refer to this model as the Arosh oscillator.

Despite the aforementioned progress, exact analytical study of the stationary state has witnessed only limited progress. The full Lindblad dynamics of a single oscillator can be explored numerically, but the exact Wigner-function evolution equation contains third-order derivatives and does not qualify as a genuine Fokker–Planck equation, which would have otherwise afforded a connection between quantum limit-cycle dynamics and an equivalent classical stochastic-process description. For coupled oscillators, the problem is even harder, because the Hilbert space grows exponentially in size with the number of oscillators, and analytical results for stationary states and synchronization measures become even more scarce.

In this paper, we first study a single quantum van der Pol oscillator as modeled by Arosh \textit{et al.}, and then exploit the insights gained in analyzing two such dissipatively-coupled oscillators. For the single oscillator, we derive the exact Wigner-function evolution equation and show that dropping the third-order terms in the equation yields a Fokker–Planck equation, which in an appropriate regime admits an equivalent It\^{o} stochastic description. We show that the stationary-state Wigner function and the corresponding radial distribution obtained from the stochastic description agree remarkably well with those from the full Lindblad master equation. We then truncate the Hilbert space to the three lowest Fock states (the qutrit subspace), wherein a parity symmetry of the Liouvillian restricts the form of the stationary state. This allows us to obtain in closed form the density matrix of the system in the stationary state, and hence the corresponding Wigner function. The latter forms a ring-like structure in phase space reminiscent of a quantum limit cycle, and displays pronounced negative regions, a signature of nonclassicality.

We next consider two non-identical Arosh oscillators with different intrinsic frequencies, which are coupled dissipatively through a common jump operator. We investigate quantum synchronization in the sense of phase locking between the two oscillators by numerically solving the full Lindblad master equation. Using the stationary-state synchronization order parameter across varying detuning and dissipative coupling regimes, we demonstrate that stronger effective nonlinear dissipation shrinks the synchronization window. We next project the oscillators onto a truncated three-level qutrit manifold. The same parity symmetry as for the single oscillator splits the nine-dimensional two-qutrit Hilbert space into an even sector (dimension 5) and an odd sector
(dimension 4), and forces the stationary-state density matrix to be block diagonal,
with no coherences between the two sectors. Consequently, only $5^2+4^2=41$ of the
$81$ matrix elements can be nonzero, which reduces the stationary-state problem to a
linear system of modest size, which we solve exactly. This tractable model matches full numerical simulations in the strong-dissipation regime while uncovering a low-dissipation signature: a split, double-peaked synchronization profile paired with an abrupt phase reversal. 

The paper is organized as follows. Section~\ref{sec:single-arosh} studies the single oscillator through the Wigner function, the Fokker–Planck equation, and the It\^{o} description, and compares them with the full quantum dynamics. Section~\ref{sec:single-arosh-qutrit} gives the exact stationary state and Wigner function in the qutrit subspace. Section~\ref{sec:two-coupled-Arosh} introduces the two coupled oscillator system and studies  synchronization in the full Hilbert space. Section~~\ref{sec:two-coupled-Arosh-qutrit} deals with the coupled system in the qutrit subspace, and Section~\ref{sec:conclusions} concludes the paper. Technical details are collected in the appendices.

\section{Single Arosh oscillator}
\label{sec:single-arosh}
We consider a quantum harmonic oscillator of unit frequency, $\omega=1$, and described by the Hamiltonian
\begin{align}
H_0= a^\dagger a,
\end{align}
with energy being measured with respect to the zero-point energy, $a^{\dagger}$ and $a$ being respectively the creation and the annihilation operator, and we work in units such that the reduced Planck's constant is unity: $\hbar=1$; we also take the mass $m$ of the oscillator to be unity. The Hilbert space of the system is spanned by the energy eigenbasis or the Fock states $\{|n\rangle\}$, with $n=0,1,2,\dots,$ and the eigenenergies are $E_{n}=n$. Following Ref.~\cite{Arosh2021}, the oscillator is taken to be interacting with a photon bath inducing photon gain and loss processes that lead to transitions between the energy eigenstates of the oscillator. The density operator evolves in time following the Lindblad master equation
\begin{align}
\dot{\rho} = -i[H_0,\rho] + \gamma_{1}\mathcal{D}[a^{\dagger}]\rho + \gamma_{2}\mathcal{D}[xa]\rho=\mathcal{L}\rho,
\label{eq:single_master}
\end{align}
where the dot denotes derivative with respect to time, $\mathcal{L}$ is the Liouvillian generator, $\gamma_{1}>0$ is the single-photon pumping rate, $\gamma_{2}>0$ is the rate associated with a nonlinear two-photon dissipative channel generated by the operator $xa$, with $x=(a+a^{\dagger})/\sqrt{2}$, and the Lindblad superoperator is
\begin{align}
\mathcal{D}[L]\rho = L\rho L^{\dagger} - \frac{1}{2}\{L^{\dagger}L,\rho\}.
\label{eq:single_master-op}
\end{align}
Note that for general values of $m$ and $\omega$, one has $x=(1/\sqrt{2m\omega})(a+a^\dagger)$. The competition between the two processes characterized by the rates $\gamma_1$ and $\gamma_2$ gives rise to a quantum limit cycle, as we will discuss below.

\subsection{Wigner-function representation}
A useful phase-space description of the quantum dynamics described by the master equation~\eqref{eq:single_master} is obtained in terms of the Wigner function. Defining the characteristic function as
\begin{align}
\chi(\beta)\equiv\mathrm{Tr}\left[\rho \exp\left(\beta a^\dagger-\beta^* a\right)\right],
\label{eq:characteristic_function}
\end{align}
with star denoting complex conjugation and Tr denoting trace operation, the Wigner function is given by
\begin{align}
W(\alpha)=\frac{1}{\pi^2}
\int_{\mathbb{C}}d^2\beta\,\chi(\beta)\exp\left(\beta\alpha^*+\beta^*\alpha\right).
\label{eq:wigner}
\end{align}
Here, the integration is over the entire complex-$\beta$ plane. The Wigner function,  normalized as $\int_{\mathbb{C}}d^2\alpha\,W(\alpha)=1$, provides a phase-space representation of the density operator, in that it is a quasiprobability distribution
function that plays a role in quantum mechanics similar
to that of the classical phase space probability distribution function. 

\begin{figure*}
\includegraphics[height=2.8cm]{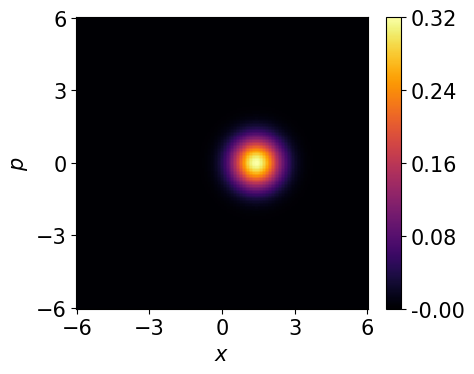}
\includegraphics[height=2.8cm]{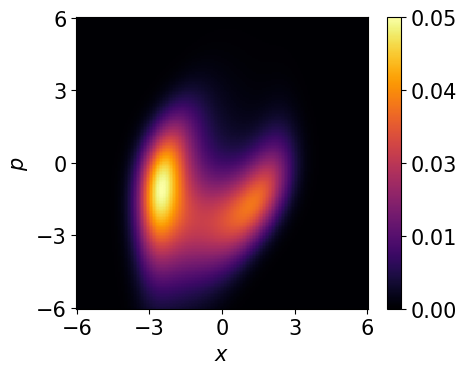}
\includegraphics[height=2.8cm]{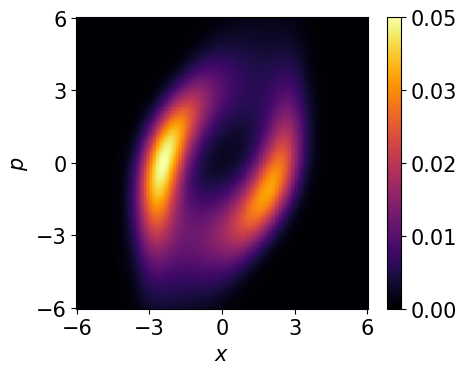}
\includegraphics[height=2.8cm]{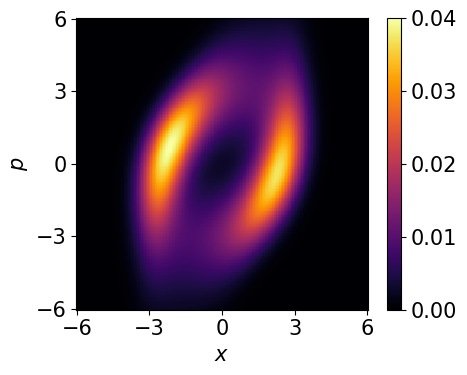}
\includegraphics[height=2.8cm]{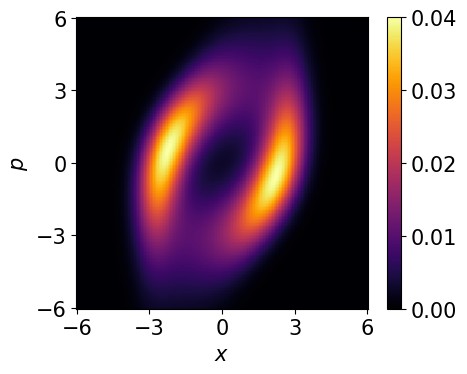}
\caption{Corresponding to the dynamics~\eqref{eq:single_master}, the figure shows the time development of the Wigner function $W_{xp}(x,p)$ starting from a coherent state $|\alpha\rangle$, with $\alpha=1.0$, and plotted in terms of the real phase-space variables $(x,p)$, defined via $\alpha=(x+ip)/\sqrt{2}$. Here, $\gamma_1=1.0$ and $\gamma_2=0.3$. The panels from left to right correspond to times $t=0$, $t=T/3$, $t=4T/3$, $t=8T/3$, and $t=3T$, respectively, where $T$ is the time period of the underlying quantum harmonic oscillator.  We observe from the plots that the Wigner function is for initial times rotating and spreading simultaneously, and then eventually reaching a stationary state in which it is spread out over a closed path that is time independent (compare the results in the last two panels). }
\label{fig: Wigner evolution van der Pol oscillator-1}
\end{figure*}

The time dependence in the Wigner function arises from that in the density operator. The time evolution while starting from a coherent state $|\alpha\rangle$ may be obtained numerically by using Eq.~\eqref{eq:wigner}, with $\rho(t)$ obtained by numerically evolving Eq.~\eqref{eq:single_master} starting with $\rho(0)=|\alpha\rangle\langle \alpha|$. In Fig.~\ref{fig: Wigner evolution van der Pol oscillator-1}, we show the Wigner function in terms of the real phase-space variables $(x,p)$, defined via 
\begin{align}
\alpha=\frac{x+ip}{\sqrt{2}}=re^{i\theta},
\end{align}
where $(r,\theta)$ define the corresponding polar coordinates. Since
$d^2\alpha=dx\,dp/2$, the Wigner function normalized
with respect to $dx\,dp$ is
\begin{align}
W_{xp}(x,p)
=\frac{1}{2}
 W\left({\alpha}\right);
\end{align} 
We have
\begin{align}
\int d^2\alpha\,W(\alpha)
=\int dx\,dp\,W_{xp}(x,p)=1.
\end{align}
We observe from Fig.~\ref{fig: Wigner evolution van der Pol oscillator-1} that the Wigner function in course of evolution attains a stationary state in which it is spread out over a time-independent closed path in the phase space. The closed path is what is referred to in the literature as a quantum limit cycle~\cite{Arosh2021}. The closed path obtained in the stationary state is independent of the choice of the initial coherent state. The stationary-state Wigner function corresponds to a time-independent density operator $\rho_\mathrm{ss}$, obtained from Eq.~\eqref{eq:single_master} by setting its left hand side (LHS) to zero. In the stationary state, evidently, the various photon loss and gain processes balance each other.  

To obtain analytically the evolution equation for $W(\alpha)$, we first transform the master equation~\eqref{eq:single_master} into an equation for the characteristic function, by using the correspondences between operator products and derivatives with respect to $\beta$ and $\beta^*$. The relevant correspondences and their derivation are given in Appendix~\ref{AppB}. The resulting equation can then be transformed into that for the Wigner function. Corresponding to the dynamics~\eqref{eq:single_master}, the exact Wigner-function evolution equation takes the form
\begin{align}
&\frac{\partial W}{\partial t}
=\Big[-\left(\frac{\partial}{\partial\alpha}\mu_\alpha+\mathrm{c.c.}\right)\nonumber \\
&+\frac{1}{2}\left(\frac{\partial^2 D_{\alpha\alpha}}{\partial\alpha^2}+\frac{\partial^2 D_{\alpha^*\alpha^*}}{\partial{\alpha^*}^2}+\frac{\partial^2 D_{\alpha\alpha^*}}{\partial\alpha\partial\alpha^*}\right)
\Big]W+\mathcal{Q}_3[W],
\label{eq:exact_wigner}
\end{align}
where c.c. stands for complex conjugate, and 
\begin{align}
\mathcal{Q}_3[W]&=\frac{\gamma_2}{16}\Big\{ \left [ \frac{\partial^3}{\partial \alpha\partial{\alpha^*}^2}[\left(-\alpha + 2\alpha^*\right)W] -
 \frac{\partial^3(\alpha W)}{\partial \alpha^3} \right] \nonumber \\
 &+\mathrm{c.c.}\Big\}
\end{align}
contains third-order derivative terms generated by the nonlinear dissipative process. Here, the drift coefficient $\mu_\alpha$ and the diffusion coefficients $D_{\alpha\alpha}$, $D_{\alpha\alpha^*}$, etc. are given by
\begin{align}
&\mu_\alpha=\left(-i+\frac{\gamma_1}{2}\right)\alpha\nonumber \\
&\qquad-\frac{\gamma_2}{4}\left(\alpha^3+\alpha^*|\alpha|^2+2\alpha|\alpha|^2-\alpha-\alpha^*\right),
\label{eq:mu_alpha}\\
&D_{\alpha\alpha}=\frac{\gamma_2}{2}\left(-\alpha^2-2|\alpha|^2+1\right),
\label{eq:D_aa}
\\
&D_{\alpha\alpha^*}=\gamma_1+\gamma_2\left(-1+\alpha^{*2}+\alpha^2+3|\alpha|^2\right),
\label{eq:D_aastar}
\end{align}
with $D_{\alpha^*\alpha^*}=D_{\alpha\alpha}^{*}$. The presence of the term $Q_3[W]$ is important: it makes Eq.~\eqref{eq:exact_wigner} unlike a genuine Fokker--Planck equation, so that the full quantum dynamics
\eqref{eq:single_master} cannot, in general, be mapped onto an equivalent classical
stochastic process. Neglecting $Q_3[W]$ leaves a second-order equation, but, as shown
below, this equation describes a bona fide diffusion process only when the diffusion
matrix is positive semidefinite, which restricts the ratio $\gamma_1/\gamma_2$.
\begin{figure}
\begin{subfigure}[b]{0.5\textwidth}
\includegraphics[width=4.43cm]{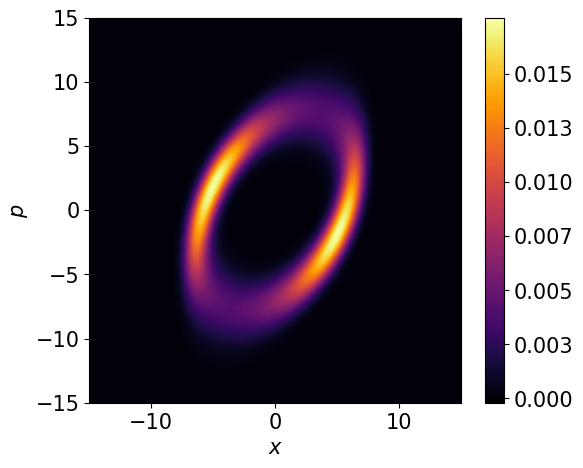} \includegraphics[width=4.43cm]{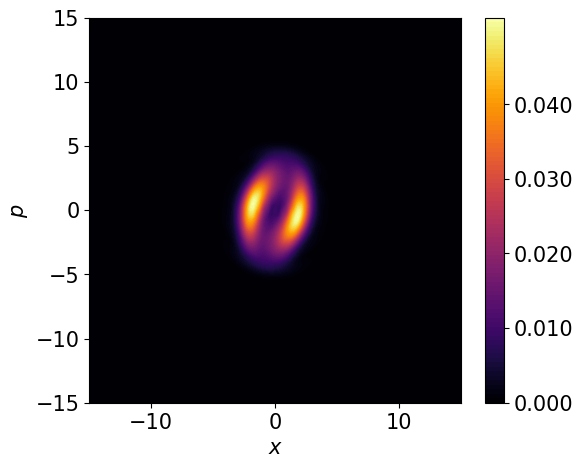}
\caption{The figure shows the Wigner function $W_{xp}(x,p)$ derived by solving numerically the master equation~\eqref{eq:single_master}, with $\gamma_1=1.0$, and $\gamma_2=0.05$ (left panel), $\gamma_2=0.5$ (right panel).}
\label{fig:WigMaster}
\end{subfigure}
\begin{subfigure}[b]{0.5\textwidth}
\includegraphics[width=4.43cm]{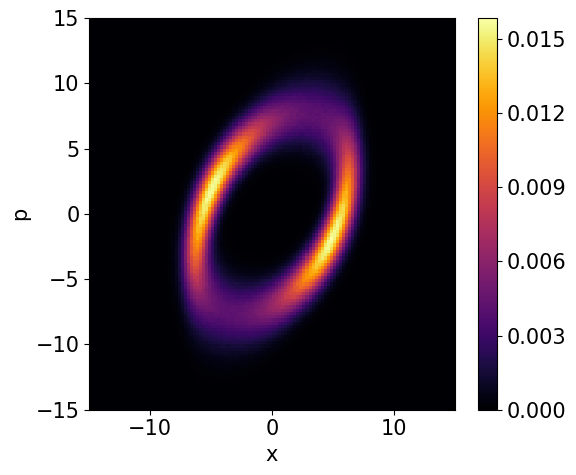} \includegraphics[width=4.43cm]{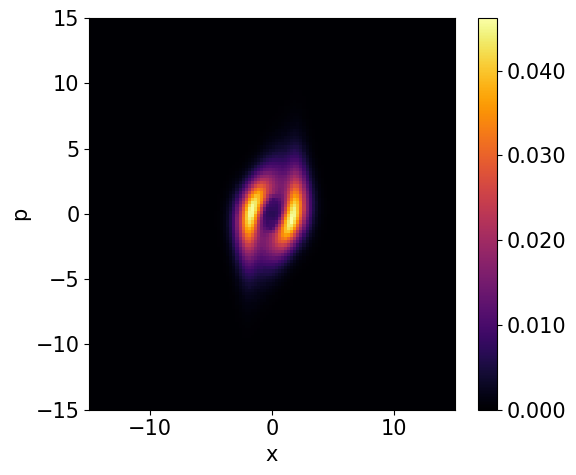}
\caption{The figure shows the Wigner function $W_{xp}(x,p)$ obtained numerically using the stochastic description~\eqref{eq:ito}, with $\gamma_1=1.0$, and $\gamma_2=0.05$ (left panel), $\gamma_2=0.5$ (right panel).}
\end{subfigure}
\caption{Comparison of the stationary-state Wigner function obtained numerically by solving the master equation~\eqref{eq:single_master} and from the stochastic description~\eqref{eq:ito}. All parameter values satisfy $\gamma_1\ge 2\gamma_2$, so that a valid stochastic description exists, see text.}
\label{fig:WigFP}
 \end{figure}
\subsection{Fokker--Planck description}
We ask: Is there an equivalent classical stochastic process that can mimic the quantum dynamics~\eqref{eq:single_master}? In order to achieve this, we neglect the term $\mathcal{Q}_3[W]$ in Eq.~\eqref{eq:exact_wigner}, to obtain the Fokker--Planck equation
\begin{align}
&\frac{\partial W}{\partial t}=\Big[-\left(\frac{\partial}{\partial\alpha}\mu_\alpha+\mathrm{c.c.}\right)\nonumber \\
&+\frac{1}{2}
\left(\frac{\partial^2 D_{\alpha\alpha}}{\partial\alpha^2}+
\frac{\partial^2 D_{\alpha^*\alpha^*}}{\partial{\alpha^*}^2}+\frac{\partial^2 D_{\alpha\alpha^*}}{\partial\alpha\partial\alpha^*}
\right)\Big]W.
\label{eq:truncated_fp_complex}
\end{align}
In terms of the phase-space variables $(x,p)$, the Fokker--Planck equation can be written in the standard two-dimensional form
\begin{align}
&\frac{\partial W_{xp}(x,p)}{\partial t}=-\frac{\partial}{\partial x}\left(\mu_x W\right)-\frac{\partial}{\partial p}
\left(\mu_p W\right)\nonumber \\
&+\frac{1}{2}\left[\frac{\partial^2}{\partial x^2}\left(D_{xx}W\right)+\frac{\partial^2}{\partial p^2}
\left(D_{pp}W\right)+2\frac{\partial^2}{\partial x\partial p}\left(D_{xp}W\right)\right],
\label{eq:real_fp}
\end{align}
with normalization $\int dxdp\,W_{xp}(x,p)=1$. The corresponding drift coefficients are
\begin{align}
   & \mu_x = \frac{\gamma_2}{2} \left(-x^3 + x \right) + \frac{\gamma_1}{2}x + p, \\ 
    &\mu_p = -\frac{\gamma_2}{2} px^2 - x +  \frac{\gamma_1}{2}p,  
    \label{eq:muy}
\end{align}
while the diffusion coefficients are
\begin{align}
    &D_{xx} = \frac{\gamma_2}{2}\left( 1  - \frac{p^2}{2} - \frac{3x^2}{2}\right) + \frac{\gamma_1}{2} \nonumber \\
    &\qquad + \frac{\gamma_2}{2}\left( -1  + \frac{p^2}{2} + \frac{5x^2}{2} \right), \label{eq:Dxx} \\ 
   & D_{pp} =  -\frac{\gamma_2}{2}\left( 1  - \frac{p^2}{2} - \frac{3x^2}{2}\right) + \frac{\gamma_1}{2} \nonumber \\
   &\qquad + \frac{\gamma_2}{2}\left( -1  + \frac{p^2}{2} + \frac{5x^2}{2} \right),  \label{eq:Dpp} \\ 
   &D_{xp} = -\frac{\gamma_2}{2} xp.\label{eq:Dxp}
\end{align}
The detailed derivation is given in Appendix~\ref{AppC}. The determinant of the diffusion matrix 
\begin{align}
D=
\begin{pmatrix}
D_{xx} & D_{xp}\\
D_{xp} & D_{pp}
\end{pmatrix}
\label{eq:diffusion_matrix}
\end{align}
is given by
\begin{align}
\det D &= \frac{\gamma_1}{2}\left(\frac{\gamma_1}{2}-\gamma_2\right)
+ \frac{\gamma_1\gamma_2}{4}p^2
+ \gamma_2\left(\frac{5\gamma_1}{4}-\frac{\gamma_2}{2}\right)x^2\nonumber \\
&+ \gamma_2^2 x^4 .
\end{align}
For $\gamma_1\ge 2\gamma_2$, all terms on the RHS are non-negative, so $D$ is positive
semidefinite throughout the phase space and the Fokker--Planck equation
\eqref{eq:real_fp} is well posed, admitting a stochastic description, as we discuss below. For $\gamma_1<2\gamma_2$, by contrast, at $x=0,p=0$, we have $D_{pp}=\gamma_1/2-\gamma_2<0$ and
$\det D<0$ in a neighbourhood of the origin; the equation then has no
stochastic interpretation. The marginal case $\gamma_1=2\gamma_2$ has
$D=\mathrm{diag}(\gamma_1/2,0)$ at $x=0,p=0$, i.e., the noise vanishes in the $p$-direction at the origin. In what follows, we
restrict to the range of values $\gamma_1\ge2\gamma_2$.

\subsection{Equivalent It\^o stochastic description}
Corresponding to the Fokker--Planck equation~\eqref{eq:real_fp}, one can write down an equivalent It\^o stochastic differential equation:
\begin{align}
d\mathbf{X}_t=\boldsymbol{\mu}(\mathbf{X}_t)\,dt+\sigma(\mathbf{X}_t)\,d\mathbf{W}_t,
\label{eq:ito}
\end{align}
where we have
\begin{align}
\mathbf{X}_t=
\begin{pmatrix}
x\\
p
\end{pmatrix},
\qquad
\boldsymbol{\mu}=
\begin{pmatrix}
\mu_x\\
\mu_p
\end{pmatrix},
\end{align}
and $\mathbf{W}_t$ is a two-dimensional Wiener process. The noise matrix $\sigma$ is related to the diffusion matrix through
\begin{align}
D=\sigma\sigma^{T},
\label{eq:sigma_relation}
\end{align}
where $T$ denotes matrix transposition. It is important to note that Eq.~\eqref{eq:ito} represents a classical stochastic process corresponding to the Fokker--Planck description~\eqref{eq:real_fp}, with the latter providing an approximate description of the full Lindblad dynamics~\eqref{eq:single_master}.

\subsection{Comparison with the full quantum dynamics}

\begin{figure}
 \centering
\includegraphics[width=\columnwidth]{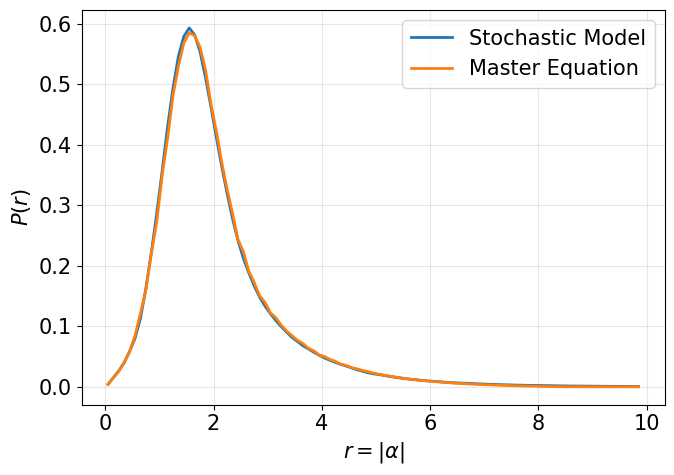}
\includegraphics[width=\columnwidth]{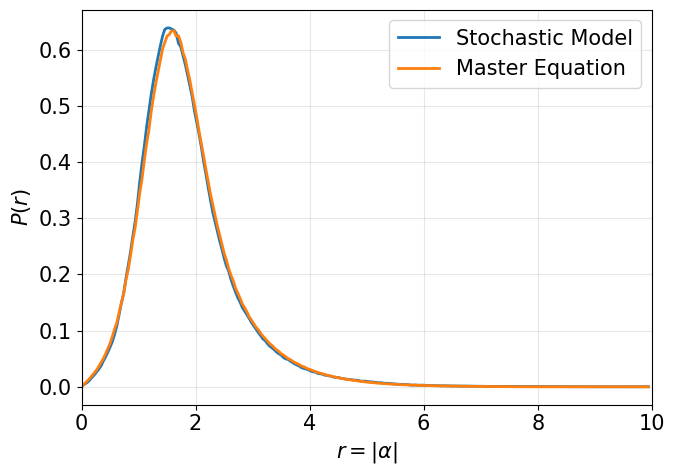} 
\includegraphics[width=\columnwidth]{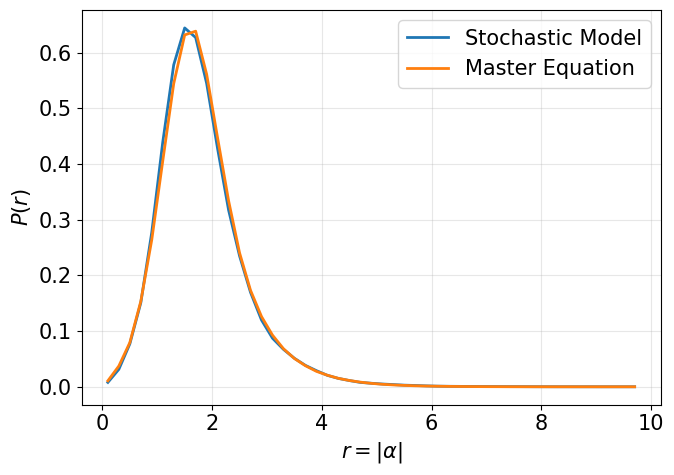}
\caption{Stationary-state radial distribution $P(r)$, obtained numerically by solving the master equation~\eqref{eq:single_master} and from the stochastic model~\eqref{eq:ito}, with $\gamma_1=2.0$, and $\gamma_2=1.0$ (top), $\gamma_1=1.0$, and $\gamma_2=0.5$ (middle), and $\gamma_1=0.8$, and $\gamma_2=0.4$ (bottom). All parameter values satisfy $\gamma_1\ge 2\gamma_2$, so that a valid stochastic description exists, see text.}
\label{fig:RadialProb}
 \end{figure} 
 
 Let us compare the approximate description given by Eq.~\eqref{eq:ito} with the dynamics obtained directly from the full Lindblad master equation~\eqref{eq:single_master}. Starting from the same initial state, the stationary-state Wigner function obtained from the two descriptions shows excellent agreement, despite the fact that higher-order derivatives are neglected in deriving the approximate description, see Figs.~\ref{fig:WigFP} and \ref{fig:RadialProb}. Thus, we may conclude that the Fokker--Planck and It\^o formulations provide useful approximate descriptions of the full dynamics~\eqref{eq:single_master}. We note that all comparisons shown here are at or inside the boundary $\gamma_1\ge 2\gamma_2$ of
validity of the stochastic description. It is noteworthy that the stochastic description gives an efficient means to mimic the system behavior in the semi-classical limit (when effectively a very large number of energy eigenstates are involved in the dynamics), which is precisely the regime where obtaining numerical results by Qutip becomes computationally very expensive. 

\section{Single Arosh oscillator in the qutrit subspace}
\label{sec:single-arosh-qutrit}
It is instructive to consider the dynamics~\eqref{eq:single_master} in the truncated Hilbert space $\mathcal{H}_{3}=\mathrm{span}\{|0\rangle,|1\rangle,|2\rangle\}$, the qutrit subspace, which affords an analytical solution for the Wigner function, while yielding at the same time insightful physical results. In this space, we have
\begin{align}
a = \begin{pmatrix} 0 & 1 & 0 \\ 0 & 0 & \sqrt{2} \\ 0 & 0 & 0 \end{pmatrix}, \quad
a^{\dagger} = \begin{pmatrix} 0 & 0 & 0 \\ 1 & 0 & 0 \\ 0 & \sqrt{2} & 0 \end{pmatrix}.
\end{align}
Note that because of truncation, we no longer have $[a,a^\dagger]=1$. Also, we have
\begin{align}
xa = \begin{pmatrix} 0 & 0 & 1 \\ 0 & \frac{1}{\sqrt{2}} & 0 \\ 0 & 0 & \sqrt{2} \end{pmatrix},
\label{eq:xa}
\end{align}
and that $(xa)^{\dagger}(xa)$ is diagonal:
\begin{align}
(xa)^{\dagger}(xa) = \begin{pmatrix} 0 & 0 & 0 \\ 0 & \frac{1}{2} & 0 \\ 0 & 0 & 3 \end{pmatrix}.
\label{eq:xadagger-xa}
\end{align}

\subsection{Parity symmetry and invariant subspace}
Consider the Fock parity operator:
\begin{align}
\Pi = \exp(i\pi a^{\dagger}a) = (-1)^{n} = \begin{pmatrix} 1 & 0 & 0 \\ 0 & -1 & 0 \\ 0 & 0 & 1 \end{pmatrix}=\Pi^\dagger,
\label{eq:Pi}
\end{align}
where $n=a^{\dagger}a$ is the number operator. Under parity transformation, we have
\begin{align}
&\Pi a^{\dagger}\Pi=-a^{\dagger}, \\
&\Pi(xa)\Pi=xa.
\end{align}
One may check that $[\Pi, H_0] = 0$. Defining the parity superoperator $\mathcal{P}_{\Pi}[\rho]\equiv\Pi\rho\Pi$, one has
\begin{align}
[\mathcal{L},\mathcal{P}_{\Pi}] = 0.
\end{align}

\subsection{Exact stationary state}
\label{sec:stationary-state}
Now, by definition, the stationary state $\rho_{\mathrm{ss}}$ of Eq.~\eqref{eq:single_master} is the null-vector of the Liouvillian generator:
\begin{align}
    \mathcal{L}\rho_{\mathrm{ss}} = 0.
    \label{eq:stationary_state_def}
\end{align}
It then follows that
\begin{align}
    \mathcal{L}(\mathcal{P}_\Pi \rho_{\mathrm{ss}}) = \mathcal{P}_\Pi(\mathcal{L}\rho_{\mathrm{ss}}) = 0.
    \label{eq:parity_commute_zero}
\end{align}
Hence, the parity-transformed state $\mathcal{P}_\Pi \rho_{\mathrm{ss}} = \Pi \rho_{\mathrm{ss}} \Pi$ is also an eigenstate of $\mathcal{L}$ with zero eigenvalue and so qualifies to be a stationary state. It is easily checked that this state has unit trace:
\begin{align}
    \mathrm{Tr}[\mathcal{P}_\Pi \rho_{\mathrm{ss}}] = \mathrm{Tr}[\Pi \rho_{\mathrm{ss}} \Pi] = \mathrm{Tr}[\Pi^2 \rho_{\mathrm{ss}}] = \mathrm{Tr}[\rho_{\mathrm{ss}}] = 1,
    \label{eq:parity_trace_unit}
\end{align}
where we have used $\Pi^2 = \mathbb{I}$. Taking Eq.~\eqref{eq:single_master} to have a unique stationary state, the unique normalization of the stationary-state density operator requires $\mathcal{P}_\Pi \rho_{\mathrm{ss}}$ to coincide identically with $\rho_{\mathrm{ss}}$:
\begin{align}
    \Pi \rho_{\mathrm{ss}} \Pi = \rho_{\mathrm{ss}}.
    \label{eq:parity_symmetry_stationary}
\end{align}
For an arbitrary $3\times3$ density matrix, this requires that
\begin{align}
\begin{pmatrix} \rho_{0,0}^{(\mathrm{ss})} & -\rho_{0,1}^{(\mathrm{ss})} & \rho_{0,2}^{(\mathrm{ss})} \\ -\rho_{1,0}^{(\mathrm{ss})} & \rho_{1,1}^{(\mathrm{ss})} & -\rho_{1,2}^{(\mathrm{ss})} \\ \rho_{2,0}^{(\mathrm{ss})} & -\rho_{2,1}^{(\mathrm{ss})} & \rho_{2,2}^{(\mathrm{ss})} \end{pmatrix} = \begin{pmatrix} 
\rho_{0,0}^{(\mathrm{ss})} & \rho_{0,1}^{(\mathrm{ss})} & \rho_{0,2}^{(\mathrm{ss})} \\ \rho_{1,0}^{(\mathrm{ss})} & \rho_{1,1}^{(\mathrm{ss})} & \rho_{1,2}^{(\mathrm{ss})} \\ \rho_{2,0}^{(\mathrm{ss})} & \rho_{2,1}^{(\mathrm{ss})} & \rho_{2,2}^{(\mathrm{ss})} \end{pmatrix},
\end{align}
which enforces
$\rho_{0,1}^{(\mathrm{ss})}=\rho_{1,0}^{(\mathrm{ss})}
=\rho_{1,2}^{(\mathrm{ss})}=\rho_{2,1}^{(\mathrm{ss})}=0$.

Proceeding, retaining only the parity-allowed matrix elements, use of Eq.~\eqref{eq:single_master} gives
\begin{widetext}
\begin{align}
\mathcal{L}\rho_{\mathrm{ss}}=
\begin{pmatrix}
-\gamma_{1}\rho_{0,0}^{(\mathrm{ss})}+\gamma_{2}\rho_{2,2}^{(\mathrm{ss})}
& 0
& -\left(\dfrac{\gamma_{1}+3\gamma_{2}}{2}-2i\right)\rho_{0,2}^{(\mathrm{ss})}
  +\sqrt{2}\gamma_{2}\rho_{2,2}^{(\mathrm{ss})}
\\
0
& \gamma_{1}\rho_{0,0}^{(\mathrm{ss})}-2\gamma_{1}\rho_{1,1}^{(\mathrm{ss})}
& 0
\\
-\left(\dfrac{\gamma_{1}+3\gamma_{2}}{2}+2i\right)\rho_{2,0}^{(\mathrm{ss})}
  +\sqrt{2}\gamma_{2}\rho_{2,2}^{(\mathrm{ss})}
& 0
& 2\gamma_{1}\rho_{1,1}^{(\mathrm{ss})}-\gamma_{2}\rho_{2,2}^{(\mathrm{ss})}
\end{pmatrix}.
\label{eq:qutrit_rhs}
\end{align}
\end{widetext}
Using 
\begin{align}
\mathcal{L}\rho_{\mathrm{ss}}=0,
\qquad
\rho_{0,0}^{(\mathrm{ss})}
+\rho_{1,1}^{(\mathrm{ss})}
+\rho_{2,2}^{(\mathrm{ss})}=1,
\end{align}
the resulting set of linear equations for the nonzero stationary-state elements of the density operator may be solved to obtain
\begin{align}
\rho_{\mathrm{ss}} =
\begin{pmatrix}
p_{0} & 0 & u_{\mathrm{ss}}+iv_{\mathrm{ss}} \\
0 & p_{1} & 0 \\
u_{\mathrm{ss}}-iv_{\mathrm{ss}} & 0 & p_{2}
\end{pmatrix},
\label{eq:rhoss}
\end{align}
with $p_n\equiv\rho_{n,n}^{(\mathrm{ss})}$, and
\begin{align}
p_{0} &= \frac{2\gamma_{2}}{2\gamma_{1}+3\gamma_{2}},
\label{eq:p0}\\
p_{1} &= \frac{\gamma_{2}}{2\gamma_{1}+3\gamma_{2}},
\label{eq:p1}\\
p_{2} &= \frac{2\gamma_{1}}{2\gamma_{1}+3\gamma_{2}},
\label{eq:p2}
\end{align}
together with 
$\rho_{0,2}^{(\mathrm{ss})}=u_{\mathrm{ss}}+iv_{\mathrm{ss}}$, where we have
\begin{align}
u_{\mathrm{ss}}
&=
\frac{4\sqrt{2}\gamma_{1}\gamma_{2}
      (\gamma_{1}+3\gamma_{2})}
     {\big[(\gamma_{1}+3\gamma_{2})^{2}+16\big]
      (2\gamma_{1}+3\gamma_{2})},
\label{eq:uvss-1}\\
v_{\mathrm{ss}}
&=
\frac{16\sqrt{2}\gamma_{1}\gamma_{2}}
     {\big[(\gamma_{1}+3\gamma_{2})^{2}+16\big]
      (2\gamma_{1}+3\gamma_{2})}.
\label{eq:uvss-2}
\end{align}
The stationary two-photon coherence can be expressed in
polar form as
$\rho_{0,2}^{(\mathrm{ss})}=R_{02}e^{i\phi_{0}}$, with  magnitude 
\begin{align}
R_{02}= \frac{4\sqrt{2}\gamma_{1}\gamma_{2}}
{(2\gamma_{1}+3\gamma_{2})
 \sqrt{(\gamma_{1}+3\gamma_{2})^{2}+16}},
\label{eq:R02}
\end{align}
and phase
\begin{align}
\phi_{0}= \arctan\left(\frac{4}{\gamma_{1}+3\gamma_{2}}\right).
\label{eq:phi0}
\end{align}

\subsection{Exact stationary-state Wigner function}
It is worth emphasizing that, upon truncation to the qutrit subspace $\mathcal{H}_3=\mathrm{span}{|0\rangle,|1\rangle,|2\rangle}$, the operators $a$ and $a^\dagger$ no longer satisfy the canonical commutation relation, $[a,a^\dagger]\neq 1$. Consequently, the standard continuous-variable Wigner-function construction is not, strictly speaking, a valid phase-space representation of the system. A genuine phase-space formulation for a finite-dimensional Hilbert space can instead be constructed using a discrete Wigner function~\cite{Wootters1987,Ferrie2011}. Nevertheless, the stationary density matrix obtained in Section~\ref{sec:stationary-state} is exact within the qutrit subspace. For the purpose of providing a phase-space visualization and connecting with the continuous-variable limit-cycle picture discussed in Section~\ref{sec:single-arosh}, we therefore evaluate this exact qutrit density matrix using the conventional oscillator Wigner-function construction. 

We now proceed to evaluate the Wigner function corresponding to the stationary-state density matrix $\rho_{\mathrm{ss}}$ at the phase-space coordinate
$\alpha=(x+ip)/\sqrt{2}=re^{i\theta}$. In terms of the displacement operator
\begin{align}
D(\alpha)=\exp(\alpha a^\dagger-\alpha^*a),
\end{align}
the Wigner function can be written as the expectation value
of the displaced parity operator:
\begin{align}
W(\alpha)
=\frac{2}{\pi}
 \mathrm{Tr}\left[\rho_{\mathrm{ss}}\Pi(\alpha)\right],
\label{eq:W_displaced_parity}
\end{align}
where
$\Pi(\alpha)=D(\alpha)\Pi D^\dagger(\alpha)$. Using $D^\dagger(\alpha)=D(-\alpha)$ and
$\Pi D(\alpha)\Pi=D(-\alpha)$ gives
$\Pi(\alpha)=\Pi D(-2\alpha)$, and hence
\begin{align}
W(\alpha)
=\frac{2}{\pi}
 \mathrm{Tr}\left[\rho_{\mathrm{ss}}\Pi D(-2\alpha)\right].
\label{eq:W_tr_form}
\end{align}

In the qutrit space, the 
matrix element $\langle m|D(-2\alpha)|n\rangle$ can be evaluated using the normal ordered expression
\begin{align}
D(-2\alpha) = e^{-2|\alpha|^2} e^{-2\alpha a^\dagger}e^{2\alpha^*a},
\label{eq:D_normal_ordered}
\end{align}
with $a|m\rangle=\sqrt{m}|m-1\rangle$, $a^\dagger|m\rangle=\sqrt{m+1}|m+1\rangle$ for $m=0,1,2$, and $a|0\rangle=0$, $a^\dagger|2\rangle=0$,
to get
\begin{widetext}
\begin{align}
D(-2\alpha)
=e^{-2|\alpha|^2}
\begin{pmatrix}
1
& 2\alpha^*
& 2\sqrt{2}(\alpha^*)^2
\\
-2\alpha
& 1-4|\alpha|^2
& 2\sqrt{2}\alpha^*(1-2|\alpha|^2)
\\
2\sqrt{2}\alpha^2
& -2\sqrt{2}\alpha(1-2|\alpha|^2)
& 1-8|\alpha|^2+8|\alpha|^4
\end{pmatrix}.
\label{eq:D_qutrit_matrix}
\end{align}
\end{widetext}
Rewriting Eq.~\eqref{eq:rhoss} as 
\begin{align}
\rho_{\mathrm{ss}}=
\begin{pmatrix}
p_0 & 0 & R_{02}e^{i\phi_0}\\
0 & p_1 & 0\\
R_{02}e^{-i\phi_0} & 0 & p_2
\end{pmatrix},
\end{align}
 using this last equation and Eqs.~\eqref{eq:Pi} and~\eqref{eq:D_qutrit_matrix} in Eq.~\eqref{eq:W_tr_form}, we get
\begin{align}
W(\alpha)
&=\frac{2}{\pi}e^{-2|\alpha|^2}
\Big[
p_0-p_1(1-4|\alpha|^2)
\nonumber\\
&\quad
+p_2(1-8|\alpha|^2+8|\alpha|^4)
\nonumber\\
&\quad
+2\sqrt{2}R_{02}
\left(\alpha^2e^{i\phi_0}
      +(\alpha^*)^2e^{-i\phi_0}\right)
\Big].
\label{eq:W_sum}
\end{align}
Substituting $\alpha=re^{i\theta}$ gives the polar
stationary-state Wigner function:
\begin{align}
W(r,\theta)
&=\frac{2}{\pi}e^{-2r^2}
\Big[
p_0-p_1+p_2
\nonumber\\
&\quad
+4(p_1-2p_2)r^2+8p_2r^4
\nonumber\\
&\quad
+4\sqrt{2}R_{02}r^2\cos(2\theta+\phi_0)
\Big].
\label{eq:wigner_arosh_full}
\end{align}
For cartesian coordinate representation, we substitute
$\alpha=(x+ip)/\sqrt{2}$. 
The cartesian expression is
\begin{align}
W_{xp}(x,p)
&=\frac{1}{\pi}e^{-(x^2+p^2)}
\Big[
p_0-p_1+p_2
\nonumber\\
&\quad
+2(p_1-2p_2)(x^2+p^2)
\nonumber\\
&\quad
+2p_2(x^2+p^2)^2
\nonumber\\
&\quad
+2\sqrt{2}R_{02}
\big((x^2-p^2)\cos\phi_0-2xp\sin\phi_0\big)
\Big].
\label{eq:W_cartesian}
\end{align}

\begin{figure*}[t]
\centering
\includegraphics[height=4.3cm]{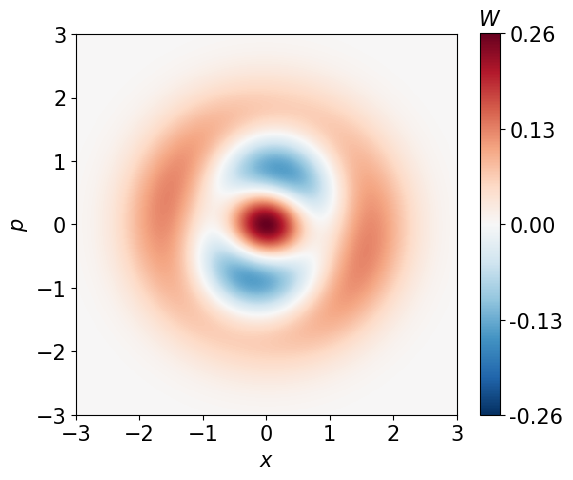}
\includegraphics[height=4.3cm]{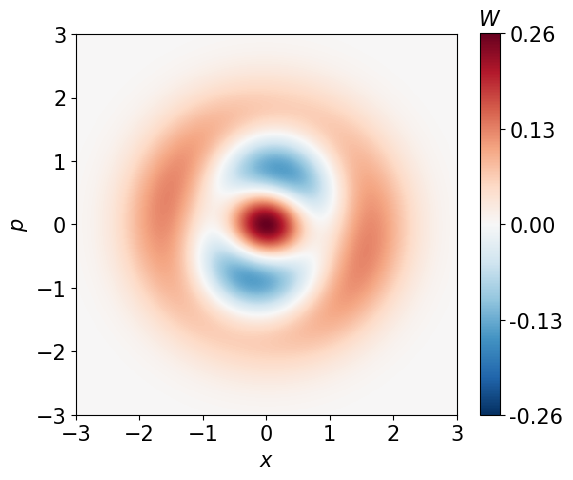}
\includegraphics[height=4.2cm]{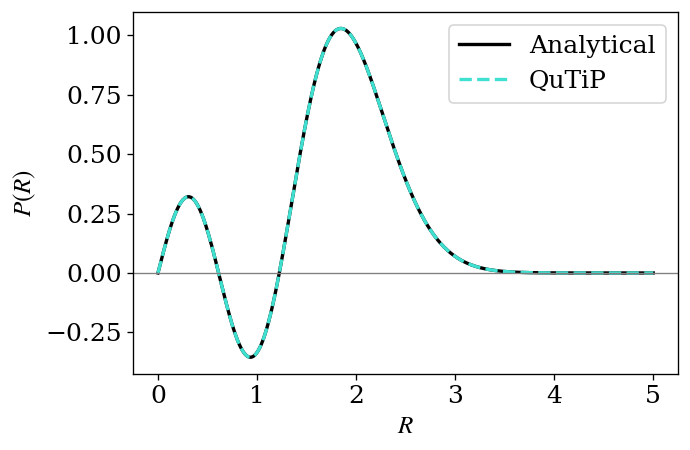}
\caption{Stationary-state Wigner function $W_{xp}(x,p)$ of the single oscillator~\eqref{eq:single_master} in the qutrit subspace, for $\gamma_1 = 4.0$, $\gamma_2 = 1.0$. The left figure is based on our analytical result, Eq.~\eqref{eq:W_cartesian}, while the middle one is based on numerical results obtained using QuTip. The figure on the right shows a comparison of the radial probability distribution obtained using the two methods, where $R = \sqrt{x^2 + p^2}$}
\label{fig:arosh_wigner-w0-6}
\end{figure*}

The Wigner function in Eq.~\eqref{eq:wigner_arosh_full}, which corresponds to the
stationary state of the Lindblad equation~\eqref{eq:single_master} projected onto
the qutrit subspace, displays a ring-like structure in phase space reminiscent of a
quantum limit cycle. Because $R_{02}\neq 0$, the Wigner function acquires an
explicit two-fold angular modulation $\cos(2\theta+\phi_{0})$, so that the ring is
not isotropic but is elongated along one axis and rotated with respect to the
coordinate axes. At the origin ($r=0$), $W(0,\theta)=(2/\pi)(p_0-p_1+p_2)=
(2/\pi)(2\gamma_1+\gamma_2)/(2\gamma_1+3\gamma_2)>0$, so that the ring surrounds a
central maximum rather than a hollow region; this feature is a consequence of the
sizeable population $p_2$ of the highest retained level and of the truncation to
$\mathcal{H}_3$. Since the qutrit stationary state, embedded in the full Fock
space, is a legitimate density operator and Eq.~\eqref{eq:wigner_arosh_full} is
its exact Wigner function, the analytical result and the numerical evaluation shown in
Fig.~\ref{fig:arosh_wigner-w0-6} agree to numerical precision, as seen both in
the phase-space plots and in the radial profile. Note that the radial profile in
Fig.~\ref{fig:arosh_wigner-w0-6} is the angular integral of $W$ and, because
$W$ can be negative, it is a quasi-probability distribution that itself takes
negative values. We emphasize that the qutrit truncation is a
strong approximation for the parameters considered here, so the above structure
should be viewed as a minimal model of the quantum limit cycle rather than a
quantitative description of the full oscillator, Eq.~\eqref{eq:single_master}.

\section{Two dissipatively-coupled Arosh oscillators: Synchronization}
\label{sec:two-coupled-Arosh}
We consider two non-identical quantum harmonic oscillators, labeled $A$ and $B$, with unit mass for each. The two oscillators are coupled dissipatively to one another. 
The dynamics of the bipartite density operator $\rho$ is governed by the Lindblad equation
\begin{align}
    &\dot{\rho} = -i [H_0, \rho] \nonumber \\
    &+ \sum_{k \in \{A, B\}} \left( \gamma_1 \mathcal{D}[L_{1,k}]\rho + \gamma_2 \mathcal{D}[L_{2,k}]\rho \right) + V \mathcal{D}[L_{\mathrm{diff}}]\rho \nonumber \\
    &=\mathcal{L}\rho,
    \label{eq:diff_master_full}
\end{align}
where the bare Hamiltonian is
\begin{align}
    H_0 = \omega_1 a^\dagger a + \omega_2 b^\dagger b.
\end{align}
Here, $\omega_1$ and $\omega_2$ are the bare frequencies of oscillators $A$ and $B$, respectively, the rates $\gamma_1$ and $\gamma_2$ are as defined for the single-oscillator case, and so are the jump operators:
\begin{align}
    L_{1,A} = a^\dagger, \quad L_{2,A} = x_A a, \qquad L_{1,B} = b^\dagger, \quad L_{2,B} = x_B b,
\end{align}
with,
\begin{align} 
    x_A = \frac{a + a^\dagger}{\sqrt{2\omega_1}}, \qquad x_B = \frac{b + b^\dagger}{\sqrt{2\omega_2}}.
\end{align}
Note that in case we consider $\omega<0$, the denominator in these expressions should consider $|\omega|$.
The coupling between the oscillators is purely dissipative,
\begin{align}
    L_{\mathrm{diff}} = a - b,
    \label{eq:L_diff_def}
\end{align}
with coupling constant $V$. Correspondingly, we have
\begin{align}
    L_{\mathrm{diff}}^\dagger L_{\mathrm{diff}} &= (a^\dagger - b^\dagger)(a - b) \nonumber \\
    &= a^\dagger a \otimes \mathbb{I}_B + \mathbb{I}_A \otimes b^\dagger b - a^\dagger \otimes b - a \otimes b^\dagger \nonumber \\
    &= n_A + n_B - (a^\dagger b + a b^\dagger),
    \label{eq:diff_norm_expansion}
\end{align}
where $n_A$ and $n_B$ are the number operators.

\begin{figure}
    \centering
    \includegraphics[width=0.49\linewidth]{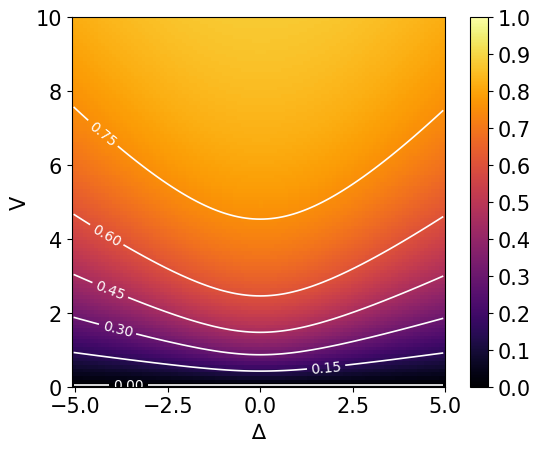}
    \includegraphics[width=0.49\linewidth]{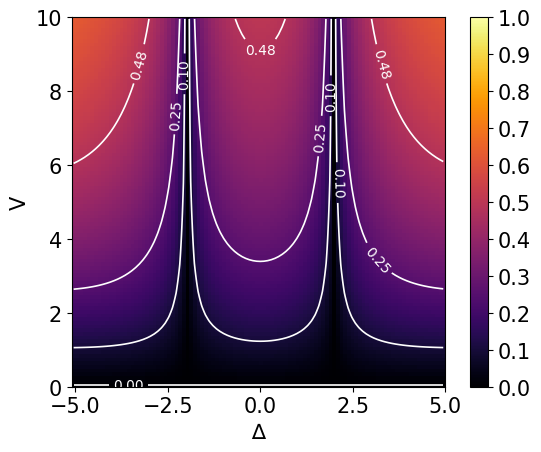}
    \includegraphics[width=0.49\linewidth]{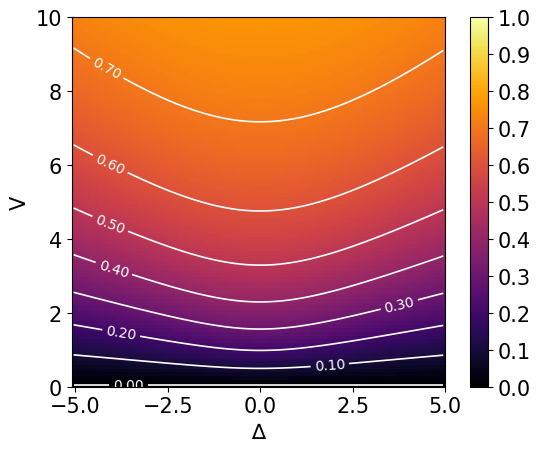}
    \includegraphics[width=0.49\linewidth]{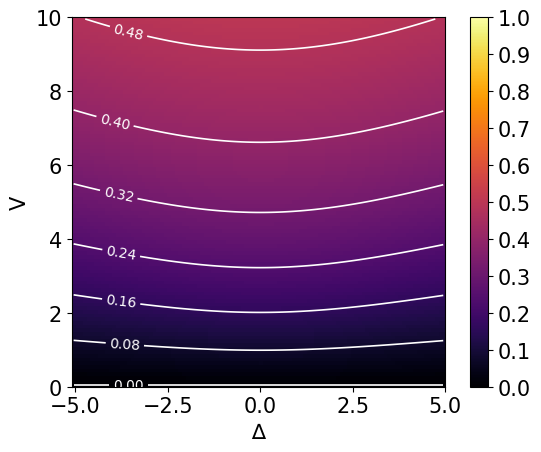}
    \caption{Stationary-state synchronization order parameter $S$ in the $(\Delta-V)$-plane, for the two coupled oscillator system, Eq.~\eqref{eq:diff_master_full}, based on numerical results obtained using QuTip. In each case, the two oscillator frequencies are taken to be $\omega_{1,2}=\omega_0\pm \Delta/2$. The parameter values are as follows. First row: $\gamma_1 =1.0$, $\gamma_2 =20.0$, with $\omega_0=10.0$ (left panel) and $\omega_0=1.0$ (right panel); Second row: $\gamma_1 =1.0$, $\omega_0=1.0$ and $\gamma_2 =5.0$ (left panel) and $\gamma_2=20.0$ (right panel). For the second row, the $x$ operator for the oscillators is taken to be frequency independent, that is, equal to $(1/\sqrt{2})(a+a^\dagger)$.}
    \label{fig:sync-full}
\end{figure}

Our interest is in synchronization between the two oscillators in the stationary state. To this end, we consider the following synchronization order parameter that measures phase locking between the two oscillators:
\begin{align}
    S = \frac{\left<a^\dagger b \right>}{\sqrt{\left<a^\dagger a\right> \left<b^\dagger b\right>}},
    \label{eq:S-defn}
\end{align}
where the quantum expectation denoted by $\langle \cdot \rangle$ is to be computed with respect to the bipartite density operator $\rho$. The absolute value $|S|$ gives the strength of phase locking, while the argument $\mathrm{arg}(S)$ gives the angle corresponding to the phase locking. The quantity $S$ is normalized such that $0\leq|S|\leq1$, whereby the value zero (respectively, the value unity) implies no synchronization (respectively, perfect synchronization). 
We evaluate numerically the quantity $S$ 
from the stationary state for from the full Liouvillian, Eq.~\eqref{eq:diff_master_full}. Figure~\ref{fig:sync-full} shows
the resulting synchronization strength in the $(\Delta,V)$ plane, where
$\Delta$ denotes the frequency mismatch. The two rows of Fig.~\ref{fig:sync-full} provide two complementary ways
of demonstrating the effect of the nonlinear dissipative rate $\gamma_2$.

In the first row, $\gamma_2$ is fixed at $\gamma_2=20$, while the mean
frequency $\omega_0$ is changed from $\omega_0=10$ in the left panel to
$\omega_0=1$ in the right panel. The oscillator frequencies are
$\omega_{1,2}=\omega_0\pm\Delta/2$, and the position quadrature entering the
nonlinear jump operator is
$x=(a+a^\dagger)/\sqrt{2|\omega|}$. Consequently, lowering $\omega_0$
increases the magnitude of the position quadrature and hence enhances the
effective strength of the nonlinear dissipative channel, even though the
bare value of $\gamma_2$ is kept unchanged. The corresponding reduction of
the region of appreciable $|S|$ in the $(\Delta,V)$ plane shows that this
increase in the effective nonlinear dissipation suppresses synchronization.
In particular, the synchronized region becomes less pronounced as the
relative importance of the nonlinear dissipative process is increased. 

One should note that the zero synchronization regions obtained in the right plot of the first row are due to $\gamma_2$ approaching $\infty$ when either of $\omega_1$ or $\omega_2$ becomes zero. The second row provides a more direct test of the same effect without the
additional frequency dependence of the position quadrature. Here we choose
the frequency-independent form
$x=(a+a^\dagger)/\sqrt{2}$ and keep $\gamma_1=1$ and $\omega_0=1$, while
increasing $\gamma_2$ directly from $\gamma_2=5$ in the left panel to
$\gamma_2=20$ in the right panel. The progressive reduction of the
synchronization region upon increasing $\gamma_2$ reproduces the trend
observed in the first row. Thus, the suppression of synchronization is not
specific to changing the oscillator frequency: it follows directly from
increasing the strength of the nonlinear dissipative channel. Figure~\ref{fig:sync-full}
therefore demonstrates that stronger nonlinear dissipation tends to
suppress phase locking between the two dissipatively-coupled oscillators.

\section{Two dissipatively-coupled Arosh oscillators in the qutrit subspace: Synchronization}
\label{sec:two-coupled-Arosh-qutrit}
Here, each oscillator is restricted to the local three-level qutrit Hilbert space $\mathcal{H}_3 = \mathrm{span}\{|0\rangle, |1\rangle, |2\rangle\}$, spanning the 9-dimensional composite Hilbert space:
\begin{align}
    \mathcal{H}_{3\times 3} = \mathcal{H}_3^A \otimes \mathcal{H}_3^B = \mathrm{span}\{|n_A, n_B\rangle\}, \quad n_A, n_B \in \{0, 1, 2\}.
\end{align}

\subsection{Parity symmetry and invariant subspaces}
We define the total Fock parity operator on $\mathcal{H}_{3\times 3}$ by
\begin{align}
    \Pi = \Pi_A \otimes \Pi_B = (-1)^{a^\dagger a + b^\dagger b} = (-1)^{n_A + n_B}.
\end{align}
One may check that
\begin{align}
    &\Pi a^\dagger \Pi= -a^\dagger, \qquad \Pi b^\dagger \Pi = -b^\dagger, \\
    &\Pi x_A \Pi= -x_A, \qquad \Pi x_B\Pi = -x_B.
\end{align}
Accordingly, for $k \in \{A, B\}$, the jump operators transform as
\begin{align}
    &\Pi L_{1,k} \Pi=-L_{1,k}, \\
    &\Pi L_{2,k} \Pi=L_{2,k}, \\
    &\Pi L_{\mathrm{diff}} \Pi=-L_{\mathrm{diff}}.
\end{align}
We then have
\begin{align}
    \mathcal{D}[-L_{\mathrm{diff}}]\rho= \mathcal{D}[L_{\mathrm{diff}}]\rho.
\end{align}
Furthermore, the bare Hamiltonian commutes with the parity operator, $[\Pi,H_0] = 0$. Defining the parity superoperator $\mathcal{P}_\Pi[\rho] \equiv \Pi \rho \Pi$, the complete Liouvillian generator $\mathcal{L}$ of Eq.~\eqref{eq:diff_master_full} satisfies
\begin{align}
    [\mathcal{L}, \mathcal{P}_\Pi] = 0.
\end{align}
Since $[\mathcal{L},\mathcal{P}_\Pi]=0$, the parity symmetry is preserved by the
dynamics: if $\Pi\rho(0)\Pi=\rho(0)$, then $\Pi\rho(t)\Pi=\rho(t)$ for all $t$.
This holds for the initial state $|0,0\rangle\langle 0,0|$. It also holds for the
stationary state: assuming $\rho_{\rm ss}$ is unique, $\mathcal{P}_\Pi\rho_{\rm ss}$ is
also a stationary state of unit trace, so that $\Pi\rho_{\rm ss}\Pi=\rho_{\rm ss}$,
as for the one-oscillator case considered earlier.

The total Fock space decomposes as $\mathcal{H}_{3\times3}=\mathcal{H}_{\rm even}\oplus
\mathcal{H}_{\rm odd}$, with
\begin{align}
\mathcal{H}_{\rm even} &= \mathrm{span}\{|0,0\rangle,|1,1\rangle,|2,0\rangle,
|0,2\rangle,|2,2\rangle\};\nonumber\\
&(n_A+n_B\in\{0,2,4\}),\\
\mathcal{H}_{\rm odd} &= \mathrm{span}\{|1,0\rangle,|0,1\rangle,|2,1\rangle,
|1,2\rangle\}, \nonumber \\
&(n_A+n_B\in\{1,3\}),
\end{align}
of dimensions $5$ and $4$, respectively. The condition $\Pi\rho\Pi=\rho$ is equivalent
to $\rho_{ij}=0$ whenever $|i\rangle$ and $|j\rangle$ belong to different sectors, so
that $\rho$ is block diagonal,
\begin{align}
\rho=\rho_{\rm ee}\oplus\rho_{\rm oo},
\label{eq:block}
\end{align}
with a $5\times5$ even--even block (25 elements) and a $4\times4$ odd--odd block
(16 elements). Note that the symmetry does \emph{not} confine the state to any given block: the jump operators $L_{1,k}$ and $L_{\rm diff}$ change the
parity of Fock states, so that both sectors are populated. What vanishes are the
coherences between the sectors. We thus conclude that all off-diagonal elements connecting states of different parities identically vanish in the stationary state ($\rho_{ij}^{(\mathrm{ss})} = 0$ for $|i\rangle \in \mathcal{H}_{\mathrm{even}}, |j\rangle \in \mathcal{H}_{\mathrm{odd}}$). The block-diagonal structure in Eq.~\eqref{eq:block} allows the stationary-state synchronization to be studied within the reduced parity-invariant subspace. 

\subsection{Exact stationary state and synchronization conditions}
\label{sec:exact-qutrit}
We now turn to solving Eq.~\eqref{eq:diff_master_full} by the method of vectorization; we will make use of the fact that the stationary-state density matrix is sparse. In the vectorized formulation, the exact $81 \times 81$ vectorized Liouvillian reads as
\begin{align}
\mathcal{L}&= -i \left( I_9 \otimes H_0 - H_0^T \otimes I_9 \right) \nonumber\\
&+ \sum_{k \in \{A, B\}} \gamma_1 \Big( L_{1,k}^* \otimes L_{1,k} - \frac{1}{2} I_9 \otimes L_{1,k}^\dagger L_{1,k} \nonumber \\
&- \frac{1}{2} (L_{1,k}^\dagger L_{1,k})^T \otimes I_9 \Big) \nonumber\\
&+ \sum_{k \in \{A, B\}} \gamma_2 \Big( L_{2,k}^* \otimes L_{2,k} - \frac{1}{2} I_9 \otimes L_{2,k}^\dagger L_{2,k} \nonumber \\
&- \frac{1}{2} (L_{2,k}^\dagger L_{2,k})^T \otimes I_9 \Big) \nonumber \\
&+ V \Big( L_{\mathrm{diff}}^* \otimes L_{\mathrm{diff}} - \frac{1}{2} I_9 \otimes L_{\mathrm{diff}}^\dagger L_{\mathrm{diff}} \nonumber \\
&- \frac{1}{2} (L_{\mathrm{diff}}^\dagger L_{\mathrm{diff}})^T \otimes I_9 \Big),
\label{eq: Vectorised Qutrit coupled}
\end{align}
where $I_9$ is the $9 \times 9$ identity matrix. Because $[\mathcal{L},\mathcal{P}_\Pi]=0$, the $41$-dimensional space of block-diagonal
operators is invariant under $\mathcal{L}$. Hence, starting from $|0,0\rangle$ (or
any parity-symmetric state), $\rho(t)$ remains block diagonal at all times.

Let $\mathrm{vec}(\rho_{\rm ss})$ be the vectorized stationary-state density matrix in
the full $81$-dimensional space, and let $x_{\rm ss}\in\mathbb{C}^{41}$ denote the vector
collecting the matrix elements $\rho_{ij}$ with $|i\rangle,|j\rangle$ in the same
parity sector (25 elements from the even--even block and 16 from the odd--odd block).
These are related via the projection matrix $P$ (of size $81\times41$, with
$P^\dagger P=I_{41}$) as
\begin{align}
\mathrm{vec}(\rho_{\rm ss})=P\,x_{\rm ss},\qquad x_{\rm ss}=P^\dagger\mathrm{vec}(\rho_{\rm ss}).
\end{align}
The dynamics in the reduced space reads $\dot x=\mathcal{L}_{\rm red}x$, with
$\mathcal{L}_{\rm red}=P^\dagger\mathcal{L}P$. This reduction is exact, since the
range of $P$ is invariant under $\mathcal{L}$, i.e.\ $\mathcal{L}P=P\mathcal{L}_{\rm red}$.

We define the trace functional in the reduced subspace as $t^\dagger = \mathrm{vec}(I_9)^\dagger P$. Enforcing the normalization condition $\mathrm{Tr}(\rho_\mathrm{ss}) = t^\dagger x_\mathrm{ss} = 1$ alongside the stationary condition $\mathcal{L}_{\mathrm{red}} x_\mathrm{ss} = 0$, we construct the non-singular matrix
\begin{align}
    M = (I_{41} - e_1 e_1^T) \mathcal{L}_{\mathrm{red}} + e_1 t^\dagger,
\end{align}
where $I_{41}$ is the $41 \times 41$ identity matrix, $e_1 = (1, 0, \dots, 0)^T \in \mathbb{R}^{41}$. The stationary-state solution satisfies $M x_\mathrm{ss} = e_1$, that is,
\begin{align}
x_\mathrm{ss} = M^{-1} e_1,
\end{align}
and the reconstructed full density matrix is $\rho_\mathrm{ss} = \mathrm{vec}^{-1}(P M^{-1} e_1)$. For any observable $A$, its stationary-state expectation value can be computed via
\begin{align}
    \langle A \rangle = \mathrm{Tr}(A \rho_\mathrm{ss}) = v_A^\dagger x_\mathrm{ss}; \quad \quad v_A^\dagger = \mathrm{vec}(A^\dagger)^\dagger P.
\end{align}

The degree of phase synchronization between the two oscillators is quantified by the normalized order parameter $S$ in Eq.~\eqref{eq:S-defn}, as
\begin{align}
    S = \frac{ v_O^\dagger M^{-1} e_1 }{\sqrt{\left( v_{n_A}^\dagger M^{-1} e_1 \right) \left( v_{n_B}^\dagger M^{-1} e_1 \right)}},
    \label{eq:S-order-parameter-qutrit}
\end{align}
where we have $O = a^\dagger b$, $n_A = a^\dagger a$, and $n_B = b^\dagger b$. Both parity blocks contribute to $S$. For example, $\langle a^\dagger a\rangle$ and
$\langle b^\dagger b\rangle$ involve the populations of the odd states
$|1,0\rangle,|0,1\rangle,|2,1\rangle,|1,2\rangle$, while $\langle a^\dagger b\rangle$
receives contributions from the odd--odd coherence $\rho_{(0,1),(1,0)}$ as well as from
even--even coherences such as $\rho_{(1,1),(2,0)}$ and $\rho_{(0,2),(1,1)}$. Since
$a^\dagger b$ commutes with $\Pi$, it has no matrix elements between the two sectors.
\begin{figure}
    \centering
    \includegraphics[width=0.49\linewidth]{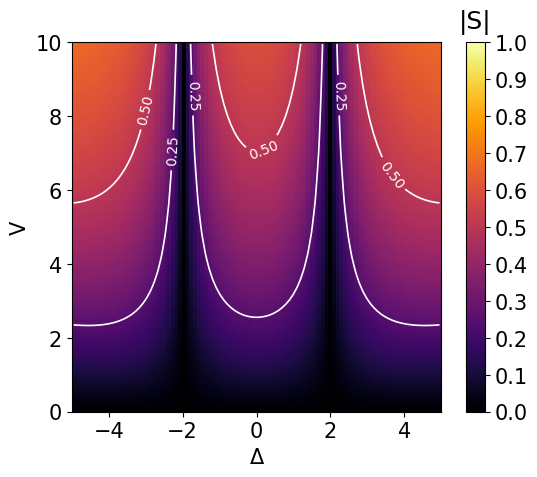}
    \includegraphics[width=0.49\linewidth]{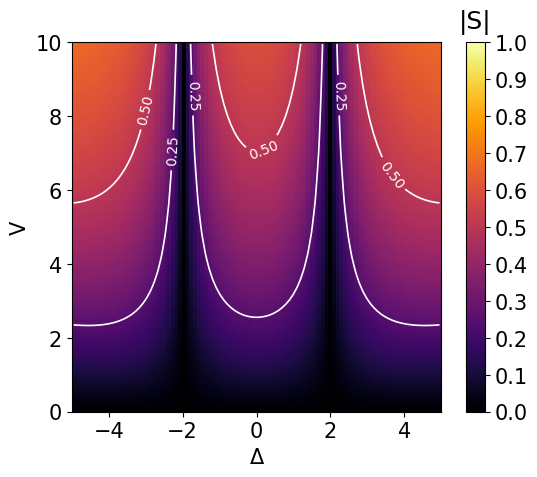}
    \includegraphics[width=0.49\linewidth]{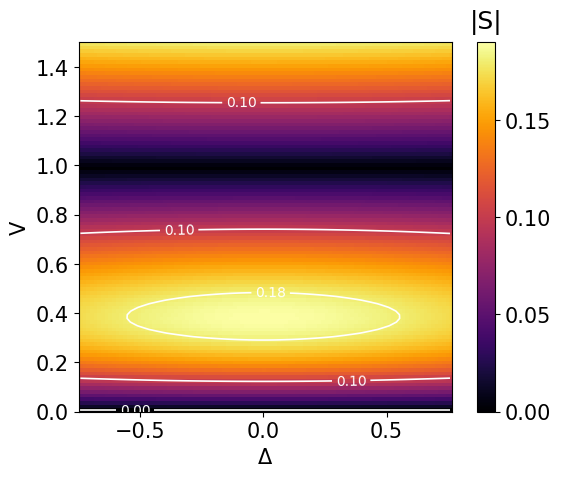}
    \includegraphics[width=0.49\linewidth]{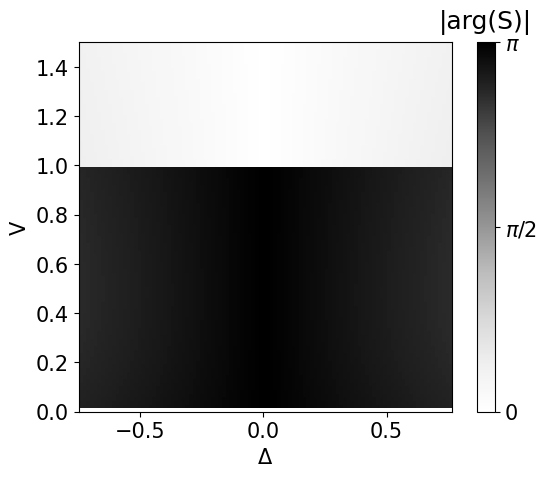}
    \caption{Stationary-state qutrit synchronization order parameter $S$ in the $(\Delta-V)$-plane, for the two coupled Qutrit system, Eq.~\eqref{eq: Vectorised Qutrit coupled}. In each case, the two oscillator frequencies are taken to be $\omega_{1,2}=\omega_0\pm \Delta/2$. The first row shows comparison between numerical results obtained using QuTip (left panel) and exact results obtained in Section~\ref{sec:exact-qutrit} (right panel) for parameter values $\gamma_1 =1.0$, $\gamma_2 =10.0$, with $\omega_0=1.0$. The second row shows the absolute value of the order parameter $|S|$ (left panel) and the corresponding phase $|\textrm{arg}(S)|$ (right panel), for parameter values $\gamma_1 =1.0$, $\gamma_2 =0.1$, with $\omega_0=10.0$. }
    \label{fig:QutritSync}
\end{figure}

Equation~\eqref{eq:S-order-parameter-qutrit} provides the synchronization order parameter directly in
terms of the inverse of the reduced $41\times41$ Liouvillian matrix $M$, and therefore allows the synchronization properties to be obtained without
constructing the full $81\times81$ stationary density matrix explicitly. The result is particularly useful for examining limiting regimes in which the structure of the synchronization becomes transparent. Figure~\ref{fig:QutritSync} illustrates two such limits, corresponding to
strong nonlinear dissipation and strong linear gain, respectively.

In the first row of Fig.~\ref{fig:QutritSync}, we consider the limit of large nonlinear dissipative rate $\gamma_2$. The left panel shows the synchronization strength $|S|$ obtained numerically from the full quantum master equation
using QuTiP, while the right panel shows the corresponding result obtained from the analytical expression in Eq.~\eqref{eq:S-order-parameter-qutrit}. The two results exhibit close agreement, demonstrating that the reduced-space construction captures the stationary-state synchronization of the full quantum model in this
regime. Moreover, the analytical result reproduces the characteristic dependence of the synchronization strength on the frequency mismatch $\Delta$ and the dissipative coupling strength $V$. In particular, the
regions of enhanced synchronization follow the same structure in the numerical and analytical calculations, providing a direct check of the reduced Liouvillian approach.

\begin{figure}
    \centering
    \includegraphics[width=1\linewidth]{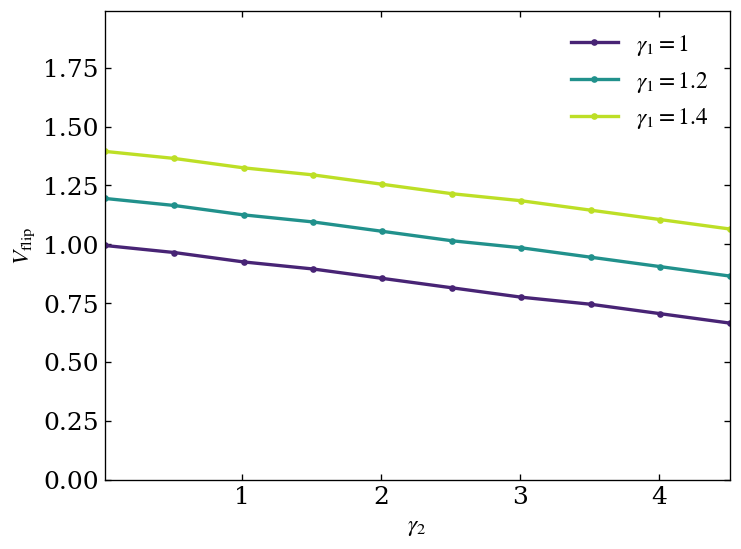}
    \caption{The coupling strength value $V_\mathrm{flip}$ at which for $\Delta=0$ a phase flip occurs (see Fig.~\ref{fig:QutritSync}, second row, left panel) is shown as a function of $\gamma_2$.}
    \label{fig:Vflip}
\end{figure}

The second row of Fig.~\ref{fig:QutritSync} considers the opposite limiting regime of large
$\gamma_1$. In this case, the synchronization strength $|S|$ develops a
pronounced double-peak structure as a function of the frequency mismatch
and coupling strength. Thus, increasing the linear gain does not simply
lead to a monotonic enhancement or suppression of synchronization; instead,
it produces two distinct regions in parameter space in which the
oscillators become more strongly phase locked. The corresponding phase
of the synchronization order parameter, $\arg(S)$, is shown in the
right panel of the second row. Interestingly, the two synchronization
regions are associated with opposite signs of the phase-locking angle,
revealing a phase reversal between the two branches. Hence, the magnitude
and phase of $S$ together show that the large-$\gamma_1$ regime supports
two distinct synchronization branches with opposite phase relations
between the oscillators. The value $V_\mathrm{flip}$ at which the phase reversal happens at $\Delta=0$ is plotted in Fig.~\ref{fig:Vflip} as a function of $\gamma_2$ and for different values of $\gamma_1$; in particular, for $\gamma_2=0$, we have $V_\mathrm{flip}=\gamma_1$. One can argue that $\gamma_2=0$, one has $|S|=0$. In this case, using 
\begin{align}
\mathcal L(I_9)
=(V-\gamma_1)\big([a,a^\dagger]+[b,b^\dagger]\big),
\end{align}
we see that at $V=\gamma_1$, the stationary state is
\begin{align}
\rho_{\rm ss}=\frac{I_9}{9},
\end{align}
which indeed yields $|S|=0$.

Our results reported above demonstrate the usefulness of Eq.~\eqref{eq:S-order-parameter-qutrit}: the reduced
parity-invariant formulation not only reproduces the synchronization
strength obtained from the full numerical calculation in the
large-$\gamma_2$ regime, but also exposes the nontrivial double-peak and
phase-reversal structure that emerges in the large-$\gamma_1$ regime.

\section{Conclusions}
\label{sec:conclusions}
We studied the quantum van der Pol oscillator and the synchronization of two such oscillators coupled dissipatively. For a single oscillator, the exact Wigner-function equation contains third-order derivatives, dropping which allows a Fokker–Planck and thence an It\^{o} stochastic-process description that agrees extremely well with the full Lindblad master equation for the system. In the qutrit subspace, parity symmetry yields the stationary state in closed form, with a Wigner function that exhibits pronounced negative regions, signaling nonclassical behavior. For the coupled system, we found that under the full Lindblad dynamics, increasing the effective nonlinear dissipative
strength suppresses the synchronized region. Restricting the two
oscillators to the qutrit subspace, a parity symmetry reduces the stationary-state problem to a small linear system, which can be solved conveniently to obtain the order parameter that quantifies phase locking and hence synchronization between the two oscillators. In this reduced description, results in the limit of large dissipation reproduce closely those from the full dynamics. On the other hand, in the limit of small dissipation, we observed a remarkable double-peak synchronization structure accompanied by a phase reversal. As future directions, one may consider exploring other symmetries that allow to obtain the stationary state in a subspace larger than the qutrit one. Another direction to pursue is study of  synchronization among a large number ($N>2$) of coupled Arosh oscillators, and exploring if in the limit $N\to \infty$ an effective mean-field description is afforded by the dynamics.

\section{Acknowledgements}
This work is supported by the Department of Atomic Energy, Government of India, under Project Identification Number RTI-4012 and RTI-4013. Computations were carried out
on the computing clusters at the Department of Theoretical Physics, TIFR, Mumbai. We would also like to thank Ajay Salve and Kapil Ghadiali for computational support. Part of this work constituted the M. Sc. thesis of VP at the Indian Institute of Science Education and Research Pune. VP acknowledges the support from the
Center for Quantum Technologies (QuTe), funded by the Transformationsfond of Saarland Federal Government. and by the QuantERA II Programme
(project “QNet: Quantum transport, metastability, and
neuromorphic applications in Quantum Networks"),
which has received funding from the EU’s Horizon
2020 research and innovation programme under Grant
Agreement No. 101017733.

\appendix

\section{Derivation of Eq.~\eqref{eq:exact_wigner}}
\label{AppB}
We start with the following definition of the Wigner function in terms of characteristic function:
\begin{align}
&W(\alpha,\alpha^*)= \frac{1}{\pi^2}\int d^2\beta~\chi^*(\beta,\beta^*)e^{\beta\alpha^*-\beta^*\alpha}; \nonumber \\
&\chi^*(\beta,\beta^*) = \mathrm{Tr}[\rho D^\dagger(\beta)],
\label{eq:Wigner-chi}
\end{align}
with $D(\beta) = e^{\beta a^\dagger - \beta^* a}$, and the integration is over the entire complex-$\beta$ plane. 

Let us first list down some useful identities. In the definition of the displacement operator, use of the Baker--Campbell--Hausdorff formula gives $D(\beta) = e^{-|\beta|^2/2} e^{\beta a^\dagger} e^{-\beta^* a}$, and taking its hermitian conjugate gives $D^\dagger(\beta) = e^{-|\beta|^2/2} e^{-\beta a^\dagger} e^{\beta^* a}$. The following identities then follow straightforwardly:
\begin{align}
&\frac{\partial D^\dagger(\beta)}{\partial \beta^*}=D^\dagger(\beta)\left(a - \frac{1}{2}\beta\right), \\ 
&\frac{\partial D^\dagger(\beta)}{\partial \beta}=-D^\dagger(\beta)\left(a^\dagger - \frac{1}{2}\beta^*\right), \\
&\frac{\partial D^\dagger(\beta)}{\partial \beta^*}= \left(a +  \frac{1}{2}\beta\right) D^\dagger(\beta), \\
&\frac{\partial D^\dagger(\beta)}{\partial \beta}=\left(-a^\dagger - \frac{1}{2}\beta^*\right) D^\dagger(\beta).
\label{eq:Dderivs-2}
\end{align}
From Eq.~\eqref{eq:Wigner-chi}, we see that $\dot{\chi}^*(\beta,\beta^*) = \mathrm{Tr}[\dot{\rho} D^\dagger(\beta)]$. Now, suppose that the Lindblad master equation had only one term: $\dot{\rho} = a\rho$. One can then derive following evolution equation for $\chi^*$:
\begin{align}
    \dot{\chi}^*(\beta,\beta^*) &= \mathrm{Tr}[a\rho D^\dagger(\beta)] = \mathrm{Tr} [D^\dagger(\beta)a\rho ] \nonumber \\
    &=\mathrm{Tr}\left[\rho\frac{\partial D^\dagger(\beta)}{\partial \beta^*}\right]+\frac{\beta}{2}\mathrm{Tr}[\rho D^\dagger(\beta)]\nonumber \\
    &=\left(\frac{\partial}{\partial \beta^*}+ \frac{\beta}{2}\right)\chi^*(\beta,\beta^*),
\end{align}
leading to the correspondence
\begin{align}
    &a\rho \rightarrow \left(\frac{\beta}{2} + \frac{\partial}{\partial\beta^*}\right).
\end{align}
Proceeding similarly, one may obtain the following correspondences: 
\begin{align}
 \label{eq:Sub Rule_app}
&\rho a \rightarrow \left(-\frac{\beta}{2}  + \frac{\partial}{\partial\beta^*}\right),\qquad a\rho  \rightarrow \left(\frac{\beta}{2}  + \frac{\partial}{\partial\beta^*}\right), \nonumber \\
&a^\dagger\rho  \rightarrow \left(\frac{\beta^*}{2}  - \frac{\partial}{\partial\beta}\right),\qquad \rho a^\dagger \rightarrow \left(-\frac{\beta^*}{2} - \frac{\partial}{\partial\beta}\right).
\end{align}

Now, consider the master equation with a generic term of the form $\dot{\rho} = abc\rho def$ where each letter can be either the creation or the annihilation operator. By the cyclic property of trace operation, it follows that it would lead to an evolution of the characteristic equation as $\dot{\chi}^*(\beta,\beta^*) = ABCFED \chi^*(\beta,\beta^*)$, where a capitalized letter denotes the respective substitution term obtained from the aforementioned correspondences. For example, if $b=a^\dagger$, then $b$ sits to the left of the density operator in the string $abc\rho def$, and so using the correspondence for $a^\dagger\rho$, we get $B = \left(\beta^*/2 - (\partial/\partial\beta)\right) $. Proceeding systematically in this manner, one obtains the evolution equation for $\chi^*(\beta,\beta^*)$. 

To obtain the evolution equation for the Wigner function, we take the Fourier transform
of the evolution equation for $\chi^*$, noting that the Wigner function is nothing but the Fourier transform of $\chi^*$, see Eq.~\eqref{eq:Wigner-chi}. Derivatives acting on $\chi^*$
are transferred to the exponential factor in the integrand by means of 
integration by parts. For the Fourier kernel
$K=e^{\beta\alpha^*-\beta^*\alpha}$, we have
$\partial K/\partial\beta=\alpha^*K$ and
$\partial K/\partial \beta^*=-\alpha K$, while multiplication by
$\beta$ and $\beta^*$ can be expressed through derivatives
with respect to $\alpha^*$ and $\alpha$, respectively. Thus, the
kernel substitutions are
\begin{align}
&\beta \longrightarrow \frac{\partial}{\partial\alpha^*},\qquad &\beta^* \longrightarrow -\frac{\partial}{\partial\alpha},\\
&\frac{\partial}{\partial\beta}\longrightarrow \alpha^*,\qquad &\frac{\partial}{\partial\beta^*}
\longrightarrow -\alpha.
\label{eq:chi-to-W-kernel}
\end{align}
In addition to these substitutions, each derivative transferred
by means of integration by parts contributes a factor of $-1$,
assuming vanishing boundary terms. Hence, a term containing
$m$ derivatives with respect to $\beta$ and $n$ derivatives
with respect to $\beta^*$ and acting on $\chi^*$ acquires the
additional factor $(-1)^{m+n}$. The order of multiplication
and differentiation operators must be preserved. Applying all of the aforementioned rules, one finally obtains Eq.~\eqref{eq:exact_wigner}.

\section{Derivation of Eq.~\eqref{eq:real_fp}}
\label{AppC}
In this appendix, we show how to systematically transform the Fokker-Planck equation~\eqref{eq:truncated_fp_complex} in complex variable $\alpha$ to the Fokker-Planck equation~\eqref{eq:real_fp} in real variables $x$ and $p$. We start with Eq.~\eqref{eq:truncated_fp_complex} and substitute $\alpha = (x + ip)/\sqrt{2}$, leading to 
\begin{align}
    &\partial_\alpha = \frac{1}{\sqrt{2}}\left(\partial_x - i\partial_p\right), \partial_{\alpha^*} = \frac{1}{\sqrt{2}}\left(\partial_x + i\partial_p\right), \\
    &\partial_\alpha\partial_{\alpha^*} = \frac{1}{2}\left(\partial^2_x + \partial_p^2\right), \\
    &\partial_{\alpha^2} = \frac{1}{2}\left(\partial^2_x - 2i\partial_{xp} - \partial^2_p\right), \\
    & \partial_{{\alpha^*}^2} = \frac{1}{2}\left(\partial^2_x + 2i\partial_{xp} - \partial^2_p\right).
\end{align}
One then obtains that 
\begin{align}
\label{Eq:coeff-real1}
   & \mu_x = \sqrt{2} \mathrm{Re}\left[\mu_\alpha\right], \\ 
    &\mu_p =  \sqrt{2}\mathrm{Im}\left[\mu_\alpha\right], \\ 
    &D_{xx} =\mathrm{Re}\left[D_{\alpha^2}\right] + \frac{1}{2} D_{\alpha\alpha^*}, \\ 
   & D_{pp} =  - \mathrm{Re}\left[D_{\alpha^2}\right] + \frac{1}{2} D_{\alpha\alpha^*}, \\ 
   &D_{xp} = \mathrm{Im}\left[D_{\alpha^2}\right].
\end{align}
Substituting $\alpha = (x + ip)/\sqrt{2}$, expanding and simplifying,  one finally obtains Eq.~\eqref{eq:real_fp}.

\bibliography{apssamp}

@article{KOVACIC2026120004,
title = {Marking 100 years of the Van der Pol equation: Overview of the dynamics of the classical Van der Pol oscillator and its generalizations},
journal = {Journal of Sound and Vibration},
volume = {643},
pages = {120004},
year = {2026},
issn = {0022-460X},
doi = {https://doi.org/10.1016/j.jsv.2026.120004},
url = {https://www.sciencedirect.com/science/article/pii/S0022460X26003652},
author = {Ivana Kovacic}
}

@misc{lahiri2026quantumsynchronizationtutorialguide,
      title={Quantum synchronization: A tutorial guide}, 
      author={Sourabh Lahiri and Shamik Gupta},
      year={2026},
      eprint={2609.14375},
      archivePrefix={arXiv},
      primaryClass={quant-ph},
      url={https://arxiv.org/abs/2609.14375}, 
}

@misc{solanki2026quantumsynchronization,
      title={Quantum Synchronization}, 
      author={Parvinder Solanki and Albert Cabot and Fernando Iemini and Federico Carollo and Midhun Krishna and Yeshma Ibrahim and Michal Hajdušek and Igor Lesanovsky and Rosario Fazio and Roberta Zambrini and Sai Vinjanampathy},
      year={2026},
      eprint={2607.19328},
      archivePrefix={arXiv},
      primaryClass={quant-ph},
      url={https://arxiv.org/abs/2607.19328}, 
}

@book{Kuramoto1984,
  title = {Chemical Oscillations,  Waves,  and Turbulence},
  ISBN = {9783642696893},
  ISSN = {0172-7389},
  url = {http://dx.doi.org/10.1007/978-3-642-69689-3},
  DOI = {10.1007/978-3-642-69689-3},
  journal = {Springer Series in Synergetics},
  publisher = {Springer Berlin Heidelberg},
  author = {Kuramoto,  Yoshiki},
  year = {1984}
}

@article{Acebron2005,
  title = {The Kuramoto model: A simple paradigm for synchronization phenomena},
  volume = {77},
  ISSN = {1539-0756},
  url = {http://dx.doi.org/10.1103/RevModPhys.77.137},
  DOI = {10.1103/revmodphys.77.137},
  number = {1},
  journal = {Rev. Mod. Phys.},
  publisher = {American Physical Society (APS)},
  author = {Acebrón,  Juan A. and Bonilla,  L. L. and Pérez Vicente,  Conrad J. and Ritort,  Félix and Spigler,  Renato},
  year = {2005},
  month = apr,
  pages = {137–185}
}

@article{nadolny2026quantum,
  author  = {Nadolny, Tobias and Bruder, Christoph},
  title   = {Quantum limit cycles and synchronization from a measurement perspective},
  journal = {Phys. Rev. Res.},
  volume  = {8},
  pages   = {023050},
  year    = {2026},
  doi     = {10.1103/c5wj-4s73}
}

@article{kehrer2025quantum,
  author  = {Kehrer, Tobias and Bruder, Christoph},
  title   = {Quantum synchronization blockade induced by nonreciprocal coupling},
  journal = {Phys. Rev. A},
  volume  = {112},
  pages   = {012223},
  year    = {2025},
  doi     = {10.1103/9y9w-lw92}
}

@article{nadolny2025nonreciprocal,
  author  = {Nadolny, Tobias and Bruder, Christoph and Brunelli, Matteo},
  title   = {Nonreciprocal Synchronization of Active Quantum Spins},
  journal = {Phys. Rev. X},
  volume  = {15},
  pages   = {011010},
  year    = {2025},
  doi     = {10.1103/PhysRevX.15.011010}
}

@article{kehrer2024quantum,
  author  = {Kehrer, Tobias and Nadolny, Tobias and Bruder, Christoph},
  title   = {Quantum synchronization through the interference blockade},
  journal = {Phys. Rev. A},
  volume  = {110},
  pages   = {042203},
  year    = {2024},
  doi     = {10.1103/PhysRevA.110.042203}
}

@article{solanki2024exotic,
  author  = {Solanki, Parvinder and Krishna, Midhun and Hajdu{\v{s}}ek, Michal
             and Bruder, Christoph and Vinjanampathy, Sai},
  title   = {Exotic Synchronization in Continuous Time Crystals Outside the Symmetric Subspace},
  journal = {Phys. Rev. Lett.},
  volume  = {133},
  pages   = {260403},
  year    = {2024},
  doi     = {10.1103/PhysRevLett.133.260403}
}

@article{nadolny2023macroscopic,
  author  = {Nadolny, Tobias and Bruder, Christoph},
  title   = {Macroscopic Quantum Synchronization Effects},
  journal = {Phys. Rev. Lett.},
  volume  = {131},
  pages   = {190402},
  year    = {2023},
  doi     = {10.1103/PhysRevLett.131.190402}
}

@article{daniel2023geometric,
  author  = {Daniel, Aaron and Bruder, Christoph and Koppenh{\"o}fer, Martin},
  title   = {Geometric Phase in Quantum Synchronization},
  journal = {Phys. Rev. Res.},
  volume  = {5},
  pages   = {023182},
  year    = {2023},
  doi     = {10.1103/PhysRevResearch.5.023182}
}

@article{koppenhoefer2019optimal,
  author  = {Koppenh{\"o}fer, Martin and Roulet, Alexandre},
  title   = {Optimal synchronization deep in the quantum regime: Resource and fundamental limit},
  journal = {Phys. Rev. A},
  volume  = {99},
  pages   = {043804},
  year    = {2019},
  doi     = {10.1103/PhysRevA.99.043804}
}

@article{roulet2018quantum,
  author  = {Roulet, Alexandre and Bruder, Christoph},
  title   = {Quantum Synchronization and Entanglement Generation},
  journal = {Phys. Rev. Lett.},
  volume  = {121},
  pages   = {063601},
  year    = {2018},
  doi     = {10.1103/PhysRevLett.121.063601}
}

@article{roulet2018synchronizing,
  author  = {Roulet, Alexandre and Bruder, Christoph},
  title   = {Synchronizing the Smallest Possible System},
  journal = {Phys. Rev. Lett.},
  volume  = {121},
  pages   = {053601},
  year    = {2018},
  doi     = {10.1103/PhysRevLett.121.053601}
}

@article{lorch2017quantum,
  author  = {L{\"o}rch, Niels and Nigg, Simon E. and Nunnenkamp, Andreas
             and Tiwari, Rakesh P. and Bruder, Christoph},
  title   = {Quantum Synchronization Blockade: Energy Quantization Hinders Synchronization of Identical Oscillators},
  journal = {Phys. Rev. Lett.},
  volume  = {118},
  pages   = {243602},
  year    = {2017},
  doi     = {10.1103/PhysRevLett.118.243602}
}

@article{lorch2016genuine,
  author  = {L{\"o}rch, Niels and Amitai, Ehud and Nunnenkamp, Andreas
             and Bruder, Christoph},
  title   = {Genuine Quantum Signatures in Synchronization of Anharmonic Self-Oscillators},
  journal = {Phys. Rev. Lett.},
  volume  = {117},
  pages   = {073601},
  year    = {2016},
  doi     = {10.1103/PhysRevLett.117.073601}
}

@article{walter2015quantum,
  author  = {Walter, Stefan and Nunnenkamp, Andreas and Bruder, Christoph},
  title   = {Quantum synchronization of two Van der Pol oscillators},
  journal = {Ann. Phys.},
  volume  = {527},
  pages   = {131--138},
  year    = {2015},
  doi     = {10.1002/andp.201400144}
}

@article{walter2014quantum,
  author  = {Walter, Stefan and Nunnenkamp, Andreas and Bruder, Christoph},
  title   = {Quantum Synchronization of a Driven Self-Sustained Oscillator},
  journal = {Phys. Rev. Lett.},
  volume  = {112},
  pages   = {094102},
  year    = {2014},
  doi     = {10.1103/PhysRevLett.112.094102}
}

@article{gupta2014kuramoto,
  author  = {Gupta, Shamik and Campa, Alessandro and Ruffo, Stefano},
  title   = {Kuramoto Model of Synchronization: Equilibrium and Nonequilibrium Aspects},
  journal = {J. Stat. Mech.},
  volume  = {2014},
  pages   = {R08001},
  year    = {2014},
  doi     = {10.1088/1742-5468/2014/08/R08001}
}

@article{rodrigues2016kuramoto,
  author  = {Rodrigues, Francisco A. and Peron, Thomas K. D. M. and Ji, Peng
             and Kurths, J{\"u}rgen},
  title   = {The Kuramoto Model in Complex Networks},
  journal = {Phys. Rep.},
  volume  = {610},
  pages   = {1--98},
  year    = {2016},
  doi     = {10.1016/j.physrep.2015.10.008}
}

@article{koppenhoefer2020quantum,
  author  = {Koppenh{\"o}fer, Martin and Bruder, Christoph and Roulet, Alexandre},
  title   = {Quantum synchronization on the {IBM Q} system},
  journal = {Phys. Rev. Res.},
  volume  = {2},
  pages   = {023026},
  year    = {2020},
  doi     = {10.1103/PhysRevResearch.2.023026}
}

@article{Vigneshwar2025,
  author  = {B. Vigneshwar and V. K. Chandrasekar and D. V. Senthilkumar},
  title   = {Dressed-state-mediated synchronization in coupled quantum oscillators},
  journal = {Phys. Rev. A},
  volume  = {112},
  pages   = {042214},
  year    = {2025},
  doi     = {10.1103/k5ry-bj1g}
}

@article{Vigneshwar2026,
  author  = {Vigneshwar, B. and Chandrasekar, V. K. and Senthilkumar, D. V.},
  title   = {Anisotropy-induced dynamics in coupled quantum oscillators},
  journal = {Phys. Rev. A},
  volume  = {113},
  pages   = {042217},
  year    = {2026},
  doi     = {10.1103/93f9-gwqv}
}

@article{Vigneshwar2026b,
  author  = {Vigneshwar, B. and Chandrasekar, V. K. and Senthilkumar, D. V.},
  title   = {Resonant blockade in driven quantum oscillators},
  journal = {Phys. Rev. A},
  volume  = {114},
  pages   = {012210},
  year    = {2026},
  doi     = {10.1103/k19y-m62n}
}

@article{Bandyopadhyay2023,
  author  = {Bandyopadhyay, Biswabibek and Banerjee, Tanmoy},
  title   = {Aging transition in coupled quantum oscillators},
  journal = {Phys. Rev. E},
  volume  = {107},
  number  = {2},
  pages   = {024204},
  year    = {2023},
  doi     = {10.1103/PhysRevE.107.024204}
}

@article{Paul2024,
  author  = {Paul, B. and Bandyopadhyay, B. and Banerjee, T.},
  title   = {Attractive-repulsive interaction in coupled quantum oscillators},
  journal = {Phys. Rev. E},
  volume  = {110},
  number  = {3},
  pages   = {034210},
  year    = {2024},
  doi     = {10.1103/PhysRevE.110.034210}
}

@article{Bandyopadhyay2021,
  author  = {Bandyopadhyay, B. and Khatun, T. and Banerjee, T.},
  title   = {Quantum Turing bifurcation: Transition from quantum amplitude death to quantum oscillation death},
  journal = {Phys. Rev. E},
  volume  = {104},
  number  = {2},
  pages   = {024214},
  year    = {2021},
  doi     = {10.1103/PhysRevE.104.024214}
}

@article{Thomas2026,
  author  = {Thomas, Nissi and Senthilvelan, M.},
  title   = {Normal-mode-splitting-induced synchronization blockade in coupled quantum van der Pol oscillators},
  journal = {Phys. Rev. A},
  volume  = {113},
  number  = {2},
  pages   = {L020202},
  year    = {2026}
}

@article{Preethi2025,
  author  = {Preethi, M. and Senthilvelan, M.},
  title   = {Quantum squeezing effects in coupled van der Pol oscillators},
  journal = {Quantum Inf. Process.},
  volume  = {24},
  number  = {4},
  pages   = {119},
  year    = {2025},
  doi     = {10.1007/s11128-025-04734-5}
}

@article{Boccaletti2002,
  author  = {Boccaletti, Stefano and Kurths, J{\"u}rgen and Osipov, Grigory and Valladares, Diego L. and Zhou, Changsong},
  title   = {The synchronization of chaotic systems},
  journal = {Phys. Rep.},
  volume  = {366},
  number  = {1--2},
  pages   = {1--101},
  year    = {2002},
  doi     = {10.1016/S0370-1573(02)00137-0}
}

@book{Gupta2018,
  title={Statistical Physics of Synchronization},
  author={Gupta, Shamik and Campa, Alessandro and Ruffo, Stefano},
  year={2018},
  publisher={Springer}
}

@article{Ferrie2011,
  author  = {Ferrie, Christopher},
  title   = {Quasi-probability representations of quantum theory with applications to quantum information science},
  journal = {Reports on Progress in Physics},
  volume  = {74},
  number  = {11},
  pages   = {116001},
  year    = {2011},
  doi     = {10.1088/0034-4885/74/11/116001}
}

@article{Wootters1987,
  author  = {Wootters, William K.},
  title   = {A Wigner-function formulation of finite-state quantum mechanics},
  journal = {Annals of Physics},
  volume  = {176},
  number  = {1},
  pages   = {1--21},
  year    = {1987},
  doi     = {10.1016/0003-4916(87)90176-X}
}

@PREAMBLE{
 "\providecommand{\noopsort}[1]{}" 
 # "\providecommand{\singleletter}[1]{#1}%" 
}

@book{strogatznonlinear,
  title={Nonlinear dynamics and chaos: with applications to physics, biology, chemistry, and engineering},
  author={Strogatz, Steven H},
  year={2018},
  publisher={CRC press}
}

@article{Arosh2021,
  title = {Quantum limit cycles and the Rayleigh and van der Pol oscillators},
  author = {Ben Arosh, Lior and Cross, M. C. and Lifshitz, Ron},
  journal = {Phys. Rev. Res.},
  volume = {3},
  issue = {1},
  pages = {013130},
  numpages = {18},
  year = {2021},
  month = {Feb},
  publisher = {American Physical Society},
  doi = {10.1103/PhysRevResearch.3.013130}
}

@book{PRK:2001, place={Cambridge}, series={Cambridge Nonlinear Science Series}, title={Synchronization: A Universal Concept in Nonlinear Sciences}, DOI={10.1017/CBO9780511755743}, publisher={Cambridge University Press}, author={Pikovsky, Arkady and Rosenblum, Michael and Kurths, Jürgen}, year={2001}, collection={Cambridge Nonlinear Science Series}}
\end{document}